\documentclass[journal=jacsat,manuscript=article]{achemso} 

\usepackage{chemformula} 
\usepackage[T1]{fontenc} 
\usepackage{booktabs}
\usepackage{multirow}
\usepackage{siunitx}
\usepackage{caption}
\usepackage{threeparttable}
\usepackage{array}
\usepackage{amsmath}
\usepackage{amsfonts}
\usepackage{hyperref}
\newcommand{\red}[1]{{\color{black} #1}}

\author{Arif Ullah}
\email{arif@ahu.edu.cn}
\affiliation[Anhui]
{School of Physics, Anhui University, Hefei, 230601, Anhui, China}
\author{Rajibul Islam}
\affiliation[Alabama]
{Department of Physics, University of Alabama at Birmingham, Birmingham 35294, AL, USA}

\author{Yangming Huang}
\affiliation[Xiamen]
{State Key Laboratory of Physical Chemistry of Solid Surfaces, College of Chemistry and Chemical Engineering, Fujian Provincial Key Laboratory of Theoretical and Computational Chemistry, and Innovation Laboratory for Sciences and Technologies of Energy Materials of Fujian Province (IKKEM), Xiamen University, Xiamen, 361005, Fujian, China}

\author{Ghulam Hussain}
\affiliation[Shenzhen]{Institute for Advanced Study, Shenzhen University, Shenzhen 518060, China}
\author{Zahir Muhammad}
\affiliation[beihang]
{National Key Laboratory of Spintronics, Hangzhou International Innovation Institute, Beihang University, Hangzhou 311115, China}
\author{Xiaoguang Li}
\affiliation[Shenzhen]{Institute for Advanced Study, Shenzhen University, Shenzhen 518060, China}
\author{Ming Yang}
\affiliation[Anhui]
{School of Physics, Anhui University, Hefei, 230601, Anhui, China}

\title[TXL Fusion for Topological Materials Discovery]
  {TXL Fusion: A Hybrid Machine Learning Framework Integrating Chemical Heuristics and Large Language Models for Topological Materials Discovery}
\abbreviations{TI, TSM, ML, LLM, DFT, XGB, SG, SOC, SHAP, PCA}

\begin{document}







\begin{abstract}
\red{Topological materials, including topological insulators (TIs) and topological semimetals (TSMs), offer promising platforms for quantum, spintronic, and low-dissipation electronic technologies. Their discovery, however, remains constrained by the high cost of first-principles calculations and the slow, resource-intensive nature of experimental validation. Here, we introduce TXL Fusion, a hybrid machine-learning framework that integrates chemically inspired heuristics, physically interpretable numerical descriptors, and large language model (LLM)-derived semantic embeddings for topological-materials classification and discovery. By combining space-group symmetry, electron-count and orbital descriptors, composition-derived topological heuristics, and physics-aware semantic representations, TXL Fusion classifies materials into trivial, TSM, and TI categories with improved overall performance and enhanced minority-class TI recognition relative to conventional descriptor-based baselines. The model further serves as a high-throughput pre-screening tool for external discovery spaces, rapidly prioritizing candidate TSMs before expensive first-principles or experimental validation. Representative TXL-prioritized candidates were subsequently supported by density functional theory (DFT) calculations, demonstrating the practical value of the framework for reducing discovery cost. By uniting symbolic chemical rules, statistical learning, and language-based representations, TXL Fusion provides a scalable and interpretable strategy for accelerating the discovery of next-generation topological and quantum materials.}
\end{abstract}

\begin{figure}[H]
    \centering
    \includegraphics[width=0.5\linewidth]{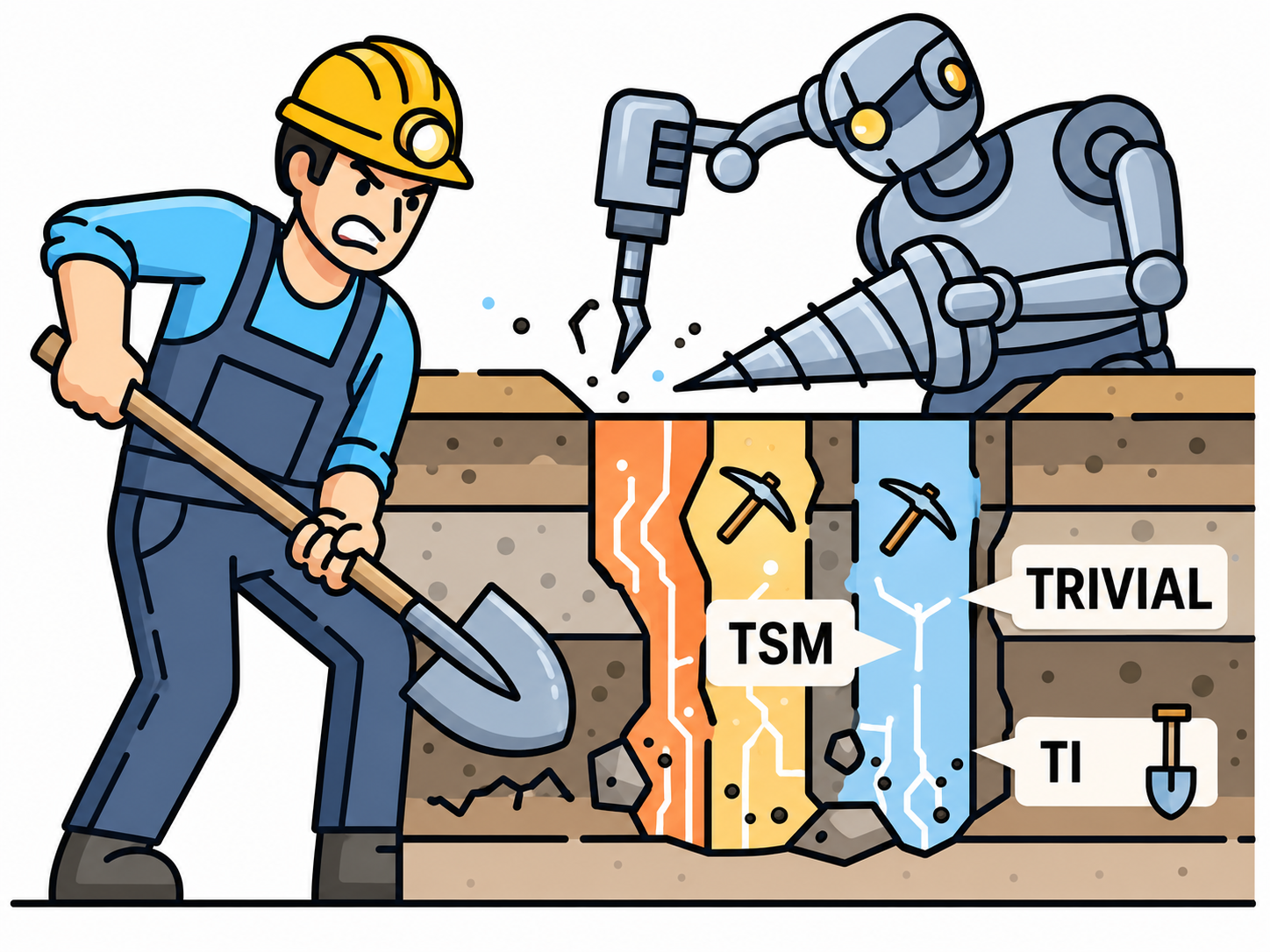}
    \caption*{}
    \label{fig:placeholder}
\end{figure}


Topological materials, encompassing topological insulators (TIs)\cite{hasan2010topological, qi2011topological} and topological semimetals (TSMs),\cite{liu2014discovery, xu2015discovery} represent unconventional quantum phases of matter characterized by nontrivial electronic band topology. Their robust boundary states, protected against perturbations such as disorder or symmetry breaking, give rise to exotic phenomena including the quantum spin Hall effect,\cite{kane2005quantum} unusual transport properties,\cite{qi2011topological} and magnetoelectric responses\cite{hu2019transport}. These unique properties position topological materials as promising candidates for next-generation quantum and spintronic technologies. Since the emergence of the field, a central challenge has been the reliable identification and classification of such materials. Early efforts relied heavily on first-principles calculations combined with topological band theory,\cite{xiao2021first, bansil2016band} a computationally intensive but powerful route for establishing topological character. The advent of symmetry indicators\cite{slager2013space, kruthoff2017topological, po2017symmetry} and topological quantum chemistry\cite{bradlyn2017topological} represented a major milestone, enabling efficient diagnosis of many topological phases directly from symmetry representations of electronic states. These symmetry-based approaches facilitated high-throughput computational searches, resulting in extensive databases of candidate topological materials and accelerating both theoretical and experimental exploration.\cite{vergniory2019complete, zhang2019catalogue, tang2019comprehensive}

Despite these advances, symmetry-based methods face inherent limitations. Certain topological phases, such as Chern insulators and time-reversal-invariant $Z_2$ insulators without point group symmetries, remain invisible to symmetry indicators and require explicit evaluation of wavefunction-based topological invariants, which is computationally expensive.\cite{po2017symmetry} Materials with low-symmetry or complex magnetic structures pose additional challenges for symmetry-based diagnosis.\cite{xu2020high} As a result, the discovery of topological materials is constrained by computational bottlenecks and the limited scope of existing frameworks.

Over the past decade, machine learning (ML) has become a scalable alternative to symmetry-based approaches for classification and property prediction of topological materials.\cite{zhang2018machine, scheurer2020unsupervised, cao2020artificial, choudhary2021high, schleder2021machine, tyner2024machine, hong2025discovery, haosheng2025predicting} Models such as gradient-boosted trees trained on space group (SG), electron count, and orbital-resolved valence descriptors have achieved strong performance,\cite{claussen2020detection} and neural networks applied to computed XANES spectra have further expanded predictive capabilities.\cite{nina2022machine} Despite these advances, conventional ML models operate solely on structured numerical inputs, limiting their ability to incorporate unstructured information—such as material descriptions, experimental annotations, or insights from scientific literature. To overcome data and scalability constraints, composition-based heuristics have been proposed, most notably the topogivity score $g(M)$,\cite{ma2023topogivity} \red{with subsequent extensions such as the integration of the Hubbard $U$ parameter for magnetic systems,\cite{xu2024discovering} the inclusion of quantum correlations across elements,\cite{xu2025quantum} and physics-informed formulations of the original heuristic.\cite{xu2026physics}} While efficient and interpretable, these composition-only rules remain insensitive to essential physical features and often struggle to distinguish closely related phases, particularly TSMs and TIs.

\red{Recent progress in large language models (LLMs) offers a complementary opportunity for materials discovery. By learning from large scientific corpora, LLMs and domain-specific transformer encoders can encode contextual chemical and physical relationships that are difficult to express as isolated numerical descriptors. Their applications now span chemistry question answering,\cite{Jablonka2024} hybrid embeddings with graph neural networks,\cite{li2025hybrid} synthesis prediction,\cite{kim2024llm} quantum-chemistry assistants,\cite{gadde2025chatbot} dataset curation,\cite{kang2025harnessing} AI-driven simulation workflows,\cite{hu2025aitomia} and crystalline-property prediction.\cite{rubungo2025llm,korolev2023accurate} Beyond direct prediction, LLM-derived embeddings can support similarity search, candidate retrieval, and multi-task learning,\cite{qu2024llm} while fine-tuning on text-encoded atomistic data has shown promise for generating physically plausible structures.\cite{gruver2024llm} These capabilities suggest that language-based representations may provide a powerful bridge between symbolic chemical knowledge, numerical descriptors, and data-driven materials classification.

However, the use of LLM-derived semantic representations for the discovery and classification of topological materials remains largely unexplored. Here, we introduce TXL Fusion, a hybrid framework that integrates three complementary sources of information: composition-driven topological heuristics, physically interpretable numerical descriptors, and semantic embeddings generated from a fine-tuned scientific language model. The framework converts structured material information into physics-aware narratives, compresses the resulting semantic embeddings into a compact representation, and combines them with heuristic and numerical descriptors through a late-fusion classifier. This design enables TXL Fusion to retain interpretability from handcrafted features while exploiting the contextual expressiveness of language-model embeddings. We show that TXL Fusion improves classification accuracy, robustness, and minority-class topological recognition relative to single-modality baselines. We further demonstrate its use as a high-throughput pre-screening tool for an external discovery space, where it rapidly prioritizes a small subset of candidates before expensive first-principles calculations or experimental validation, thereby reducing the computational and experimental burden of materials discovery. Several TXL-prioritized candidates were further supported by first-principles electronic-structure analysis as TSM candidates. Our results establish semantic--numerical fusion as a scalable strategy for topological-materials discovery, demonstrating how symbolic chemical rules, statistical learning, and language-based representations can be integrated to address complex materials-discovery challenges.}

\section{Dataset and Feature Selection} \label{sec:data-analysis-main}

\red{We source our data from the topological materials database,\cite{topo_materials, bilbao_cryst, bradlyn2017topological, vergniory2019complete, vergniory2022all} which includes DFT calculations with spin--orbit coupling (SOC). After removing formula-plus-SG entries with contradictory class labels (see Section~S11 of the Supporting Information), the cleaned dataset used in this work contains 36,953 materials, comprising 5,587 TIs ($\sim$15.1\%), 13,652 TSMs ($\sim$36.9\%), and 17,714 trivial materials ($\sim$47.9\%).} Guided by both theoretical considerations and systematic empirical analysis, we conducted a comprehensive feature selection process; our initial feature set spanned many properties including chemical bonding characteristics (e.g., covalent vs.\ ionic tendencies), SOC strength ($\propto Z^4$), periodic table group and period positions, total number of electrons, SG, valence electrons, and atomic mass. Through iterative evaluation, we refined this broad feature pool to a compact set of descriptors that consistently offered both statistical robustness and physical interpretability. Further methodological details and extended analysis are provided in Section~S1 of the Supporting Information.

Based on our analyses, SG symmetry emerged as one of the most decisive indicators of electronic class. In the cleaned dataset, 216 unique SGs are represented. Trivial compounds are most frequently found in SGs 14, 62, 2, and 15, while TSMs are concentrated in high-symmetry SGs such as 225, 194, 221, and 139. TIs most frequently occur in SGs 62, 63, 139, and 12, indicating that symmetry strongly constrains topological behavior but is not sufficient by itself to determine class membership. Several SGs are entirely absent in specific classes, confirming strong symmetry selectivity across topological phases (Supporting Information, Fig.~S1 and Table~S1).

Complementary chemical and electronic descriptors further improve class separability. TIs and TSMs are strongly enriched in transition-metal and lanthanide content, consistent with enhanced SOC and increased likelihood of band inversion. By contrast, trivial materials contain a substantially larger nonmetal fraction and higher average $p$-orbital occupation, indicative of more localized bonding with weaker orbital hybridization. Electron-count parity provides an additional discriminative signal: 72.0\% of TSMs possess odd electron counts, consistent with gapless or partially filled electronic structures, whereas most TIs (85.5\%) and trivial materials (96.0\%) exhibit even electron counts compatible with full band filling. Bonding analysis further shows that trivial compounds are predominantly moderately ionic (58.9\%), whereas TIs and TSMs are primarily mostly covalent (56.8\% and 55.1\%, respectively). This trend suggests that nontrivial topology is favored by more delocalized bonding and stronger orbital hybridization, which facilitate band inversion and topological phase formation (Supporting Information, Table~S2).

Collectively, these insights establish a concise and interpretable feature space—integrating symmetry, orbital, compositional, and bonding descriptors—that forms the foundation of our TXL Fusion framework presented below.

\begin{figure*}
    \centering
    \includegraphics[width=1.0\linewidth]{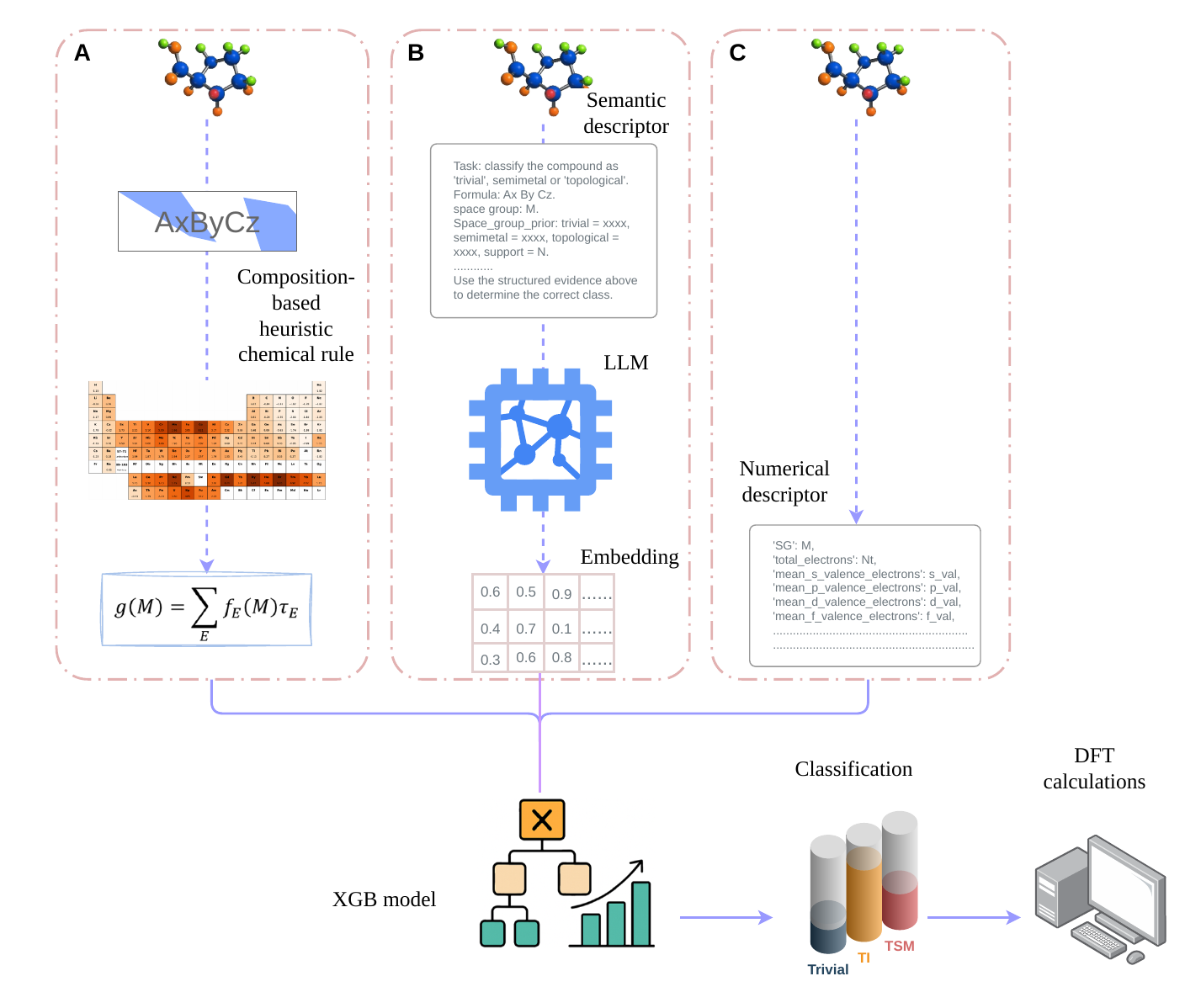}
    \caption{Schematic flowchart of the TXL Fusion model, outlining the main stages of the workflow.}
    \label{fig:flowchart}
\end{figure*}

\section{TXL Fusion architecture}
The TXL Fusion model integrates chemically inspired heuristics, numerical descriptors, and LLM embeddings within a unified hybrid framework that couples domain intuition with data-driven learning for robust classification of topological materials. As illustrated in Fig.~\ref{fig:flowchart}, the framework consists of three interconnected modules, each detailed in the Supporting Information (Section S2).

The pipeline begins with a composition-based heuristic module, adapted from 
Ma et al.\cite{ma2023topogivity}, that assigns elemental contribution scores 
to estimate the likelihood of a material belonging to the trivial, TSM, or TI 
class. This heuristic captures chemically intuitive trends—lighter, nonmetallic 
elements favour trivial phases, while heavier elements such as Bi, Sb, and Te 
correlate with topological behaviour—yet its discriminative power is limited, 
particularly for distinguishing the closely related TI and TSM phases 
(Fig.~\ref{fig:ablation_comparison_bars} and Results and Discussion). To 
overcome this, we complement the heuristic with a numerical descriptor module 
that explicitly encodes symmetry, electron‑count, orbital, bonding, and 
compositional information.

The numerical descriptor module encodes physically meaningful quantities—space
group symmetry, total and parity‑resolved electron counts, orbital occupancies,
electronegativity differences, and compositional ratios—into a fixed‑length
vector. Selected through the feature analysis described in Section ``Dataset and
Feature Selection'' and Supporting Information Section~S1, these descriptors
provide a systematic, interpretable representation of the material's electronic
and structural properties.

The third component in our pipeline is the LLM embedding module, built upon a fine-tuned SciBERT encoder, which converts structured textual descriptions of materials (including chemical formulas, SG annotations, orbital contributions, and heuristic-derived reasoning) into dense semantic embeddings. These embeddings capture contextual and higher-order correlations beyond what explicit numerical features represent, linking symbolic chemical knowledge with statistical learning.
\red{
Finally, the heuristic outputs, curated numerical descriptors, and LLM-derived semantic embeddings are integrated through a late-fusion framework. The 25-dimensional numerical--heuristic feature block is combined with a principal component analysis (PCA)-compressed 50-dimensional semantic embedding to form a compact fused representation for each material. A detailed analysis of the number of retained PCA components and its effect on model performance is provided in Supporting Information Section~S4. Rather than relying on simple concatenation alone, TXL Fusion employs a learned gating mechanism to adaptively weight semantic and numerical contributions on a per-sample basis, followed by a hierarchical two-stage eXtreme Gradient Boosting (XGB) refinement that first separates trivial from nontrivial materials and then discriminates TSMs from TIs. Final class probabilities are further calibrated using validation-optimized blending and class-specific decision thresholds. This multi-level fusion strategy enables TXL Fusion to combine interpretability with predictive performance, yielding a scalable and generalizable framework for the intelligent discovery of topological materials. Detailed implementation procedures, PCA-component sensitivity analysis, and model specifications are provided in the Supporting Information (Sections~S2--S4).}

\section{Results and Discussion}
\label{sec:results-discussion}
\red{
We evaluate TXL Fusion by comparing it with the standalone numerical-descriptor XGB baseline and with a sequence of ablated models that isolate the heuristic, numerical, and LLM-derived semantic branches. Detailed training procedures are provided in the Supporting Information (Section~S3). After data cleaning, the dataset was divided into a 29,556-material training pool (80.0\%) and a fixed 7,397-material held-out test set (20.0\%). The training pool was further split into 23,644 subtraining materials (80.0\% of the training pool; 64.0\% of the cleaned dataset) and 5,912 validation materials (20.0\% of the training pool; 16.0\% of the cleaned dataset). This protocol ensures that model selection, hyperparameter tuning, fusion calibration, and threshold optimization were performed exclusively on the subtraining and validation splits, while the held-out test set remained untouched until final evaluation.

Because the cleaned dataset is class-imbalanced, with TIs forming the minority class, unweighted training can bias the models toward the majority trivial and TSM classes. We therefore first examined the effect of class weighting on both standalone XGB and TXL Fusion, comparing unweighted training, inverse-frequency balancing, and stronger TI-specific up-weighting (Supporting Information Section~S7). This analysis showed that inverse-frequency balancing provides the most stable compromise: it improves TI recovery relative to unweighted training while avoiding the reduction in global accuracy and macro-F1 caused by aggressive TI over-weighting. We therefore adopted balanced class weighting for all performance comparisons reported below.

To place all model components under a common evaluation protocol, we also report the validation and held-out performance of the direct three-class extension of the composition-based heuristic $g(M)$. All models were assessed using precision, recall, and F1‑score:
\begin{equation}
\text{Precision} = \frac{TP}{TP + FP}, \quad
\text{Recall} = \frac{TP}{TP + FN}, \quad
\text{F1} = 2 \cdot \frac{\text{Precision} \cdot \text{Recall}}{\text{Precision} + \text{Recall}} ,
\end{equation}
where $TP$, $FP$, and $FN$ denote true positives, false positives, and false negatives, respectively. Precision measures the fraction of predicted positives that are correct, recall measures the fraction of true positives recovered, and F1-score provides their harmonic mean.

\subsection{Ablation analysis of semantic--numerical fusion}
\label{sec:ablation_analysis}

To quantify the contribution of each component in TXL Fusion, we compared five progressively enriched models: the composition-based heuristic rule ($g(M)$), the standalone XGB model trained on numerical descriptors (XGB), heuristic-enhanced XGB ($g(M)$ + XGB), Heuristic+LLM ($g(M)$ + LLM), and the final TXL Fusion model (Fig.~\ref{fig:ablation_comparison_bars}). Here, Heuristic+LLM denotes an XGB classifier trained on the 50 PCA-compressed semantic components derived from the fine-tuned SciBERT narratives, in which the heuristic $g(M)$ values are incorporated into the structured material descriptions. This model therefore evaluates the contribution of the semantic descriptor branch used in TXL Fusion, including embedded heuristic reasoning. Asterisks in Fig.~\ref{fig:ablation_comparison_bars} indicate the highest-performing model for each metric within the corresponding panel.

The overall metrics show a clear and reproducible performance hierarchy on both validation and held-out test sets. The heuristic-only model provides a useful but limited baseline, with validation accuracy, macro-F1, and weighted-F1 values of 0.686, 0.597, and 0.679, and corresponding held-out values of 0.678, 0.595, and 0.669. Replacing this rule-based model with standalone numerical XGB produces a large improvement, increasing the validation metrics to 0.828, 0.787, and 0.836, and the held-out metrics to 0.820, 0.783, and 0.828. This confirms that symmetry, electron-count, orbital, bonding, and compositional descriptors provide substantially richer predictive information than global composition-based trends alone. Adding explicit heuristic information to XGB gives a further but modest gain, with validation metrics of 0.837, 0.797, and 0.844, and held-out metrics of 0.825, 0.788, and 0.832. A larger improvement is obtained by Heuristic+LLM, which reaches 0.845, 0.807, and 0.852 on validation and 0.841, 0.808, and 0.848 on the held-out test set, demonstrating that the semantic components carry substantial predictive information beyond handcrafted descriptors. The final TXL Fusion model achieves the strongest overall performance on both splits, with validation accuracy, macro-F1, and weighted-F1 values of 0.869, 0.827, and 0.870, and held-out values of 0.858, 0.822, and 0.860. Thus, the global comparison follows the same ordering on validation and held-out data: Heuristic-only $<$ Numerical XGB $<$ Heuristic-enhanced XGB $<$ Heuristic+LLM $<$ TXL Fusion.

The class-wise held-out results explain how these global gains arise. For trivial materials, the heuristic-only model already attains relatively high recall (0.877) but limited precision (0.763), yielding an F1-score of 0.816 (Fig.~\ref{fig:ablation_comparison_bars}D). Numerical XGB substantially improves the trivial F1 to 0.884, and heuristic-enhanced XGB gives a small additional increase to 0.886. Heuristic+LLM further improves trivial F1 to 0.899 and gives the highest trivial precision among all models (0.934), indicating that the semantic descriptor is effective at identifying high-confidence trivial compounds. TXL Fusion slightly reduces precision relative to Heuristic+LLM (0.923 versus 0.934), but improves recall from 0.866 to 0.894, producing the best held-out trivial F1 of 0.908. Thus, for trivial compounds, TXL improves the final F1 primarily by recovering more true trivial samples while maintaining strong precision.

For TSMs, the heuristic-only model performs poorly, with held-out precision, recall, and F1-score values of 0.679, 0.567, and 0.618, respectively (Fig.~\ref{fig:ablation_comparison_bars}F). This confirms that composition-level trends alone are insufficient for reliable TSM identification. Numerical XGB improves TSM F1 to 0.846, and heuristic-enhanced XGB gives a modest increase to 0.850. Heuristic+LLM improves TSM F1 further to 0.862 and achieves the highest TSM precision (0.887), again showing that the semantic descriptor captures high-confidence class-specific patterns. TXL Fusion gives the best held-out TSM F1 of 0.875 by increasing recall from 0.839 for Heuristic+LLM to 0.873, while maintaining competitive precision (0.876). Therefore, for TSMs, the final fusion step improves the precision--recall balance mainly by recovering more true TSM compounds.

The minority TI class exhibits a different trade-off. The heuristic-only model has limited TI performance, with a held-out F1-score of 0.351 (Fig.~\ref{fig:ablation_comparison_bars}H). Numerical XGB improves TI F1 to 0.621, and heuristic-enhanced XGB increases it slightly to 0.629. Heuristic+LLM gives a larger improvement, raising TI F1 to 0.662 and achieving the highest TI recall among all models (0.774). This indicates that the semantic descriptor is particularly effective at recovering true TI compounds. However, this high recall is accompanied by moderate precision (0.578). In contrast, TXL Fusion lowers TI recall to 0.718 but substantially increases TI precision to 0.654, the highest among all compared models. This precision gain raises TI F1 to 0.684, giving the best TI performance overall. Thus, TXL does not simply predict more compounds as topological; rather, it improves the reliability of TI assignments by reducing false-positive TI predictions.

The validation results show the same class-wise trend (Fig.~\ref{fig:ablation_comparison_bars}C, E, G). TXL Fusion gives the highest validation F1 for all three classes, reaching 0.917 for trivial materials, 0.888 for TSMs, and 0.676 for TIs. For trivial and TSM classes, TXL improves F1 mainly through higher recall relative to Heuristic+LLM, whereas for TIs it achieves the highest precision and F1 despite lower recall than Heuristic+LLM. The agreement between validation and held-out trends indicates that the observed gains are not specific to the final test split.

Taken together, the ablation results show that each model component plays a distinct role. The heuristic rule captures broad chemical trends but lacks sufficient discriminative power as a standalone classifier. Numerical XGB provides a strong and physically interpretable baseline. Heuristic-enhanced XGB shows that explicit heuristic scores add weak but useful chemical prior information. Heuristic+LLM demonstrates that PCA-compressed semantic descriptors encode powerful contextual information, improving overall performance and providing strong class-specific precision or recall. The final TXL Fusion model gives the best global metrics and the best F1-score for all three classes by integrating semantic, numerical, and heuristic evidence into a more balanced classifier. This staged improvement confirms that the performance gain of TXL Fusion arises from the complementary interaction between semantic context and physically interpretable numerical information, rather than from any single descriptor type alone.

\begin{figure*}[htbp]
\centering
\includegraphics[width=0.90\linewidth]{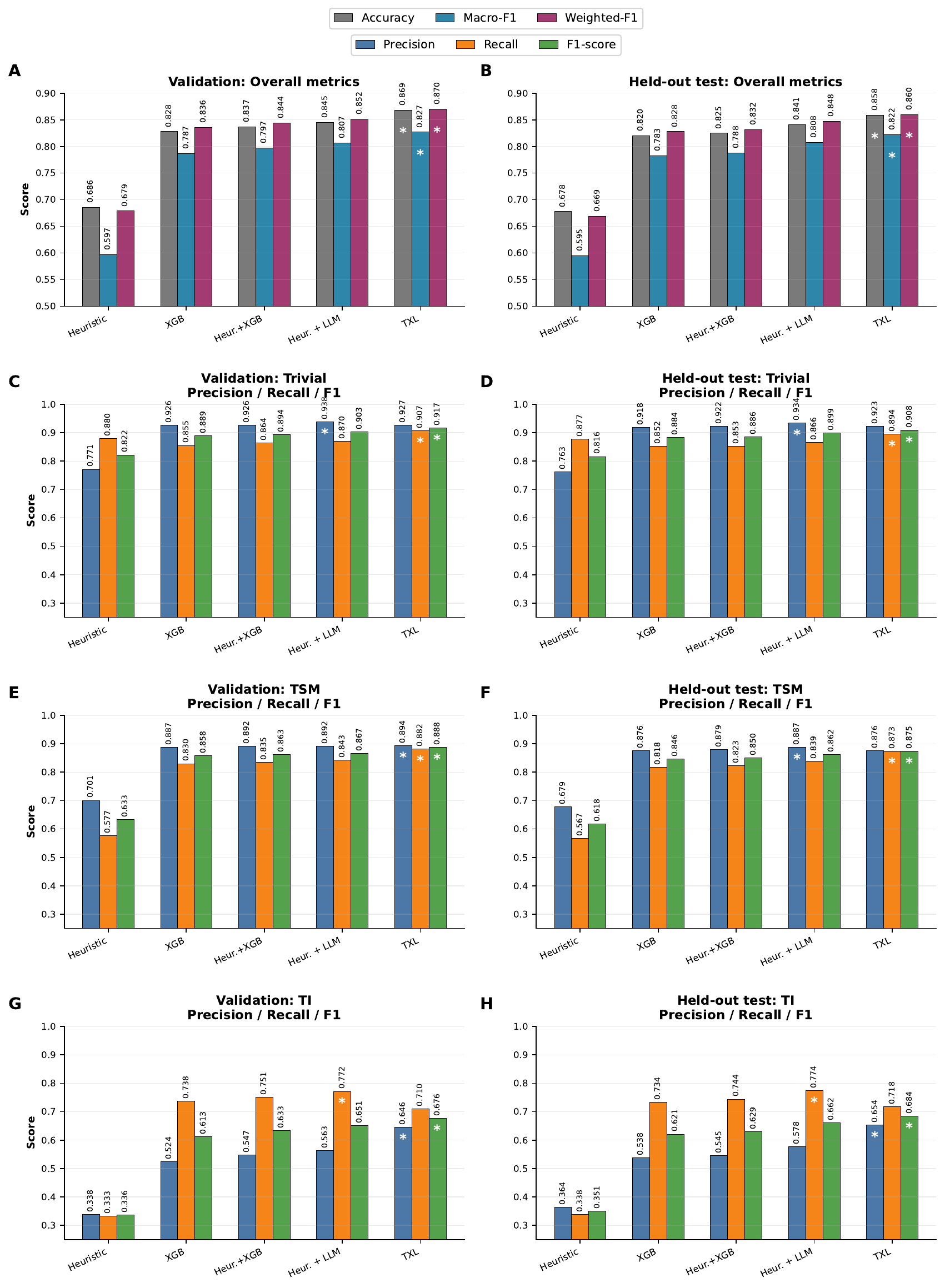}
\caption{\red{Ablation comparison of TXL Fusion. Panels A and B report overall validation and held-out test metrics, respectively, including accuracy, macro-F1, and weighted-F1. Panels C--H report class-wise validation and held-out precision, recall, and F1-score for trivial materials, TSMs, and TIs. Asterisks indicate the highest-performing model for each metric within the corresponding panel.}}
\label{fig:ablation_comparison_bars}
\end{figure*}

\subsection{Element-resolved performance and model calibration}
\label{sec:element_calibration_main}

Element-resolved evaluation shows that TXL Fusion outperforms the standalone XGB baseline across nearly all composition-complexity regimes, with the most consistent gains observed for the minority TI class (Supporting Information Sections~S6.1--S6.3). Across compounds containing one to six constituent elements, TXL improves TI F1 in every bin: unary ($0.29 \to 0.39$), binary ($0.62 \to 0.66$), ternary ($0.66 \to 0.73$), quaternary ($0.50 \to 0.59$), five-element ($0.44 \to 0.50$), and six-element compounds ($0.33 \to 0.40$). TSM performance also improves in most regimes, including unary ($0.75 \to 0.79$), binary ($0.83 \to 0.86$), ternary ($0.86 \to 0.89$), five-element ($0.87 \to 0.93$), and six-element materials ($0.83 \to 0.91$), with only a marginal decrease for quaternary TSMs ($0.84 \to 0.83$). Trivial compounds remain robust, with improvements in unary ($0.79 \to 0.89$) and six-element systems ($0.94 \to 0.97$), while five-element trivial performance remains unchanged at 0.97 for both models. These results show that the advantage of TXL Fusion is not confined to the aggregate test set but persists across chemically simple and chemically complex regimes.

The gains are especially informative in low-support regions of chemical space. Unary compounds account for only 0.72\%, 1.83\%, and 1.41\% of the trivial, TSM, and TI classes, respectively; six-element systems are also rare across all classes (1.47\%, 0.25\%, and 0.13\%); and five-element compounds are sparse for TSMs and TIs. Despite this limited support, TXL improves held-out F1 in nearly all sparse subsets. However, the magnitude of improvement remains smaller for unary materials than for multicomponent systems. In unary TIs, TXL raises F1 from 0.29 to 0.39, but this remains far below the performance achieved for binary, ternary, and quaternary TIs. A dedicated analysis of low-complexity materials in Section~S6.2 (Supporting Information) attributes this difficulty to descriptor degeneracy: when only one or two atomic species are present, many composition-based descriptors become constant or weakly varying, forcing the model to rely on indirect symmetry, orbital, and electron-count proxies. By contrast, multicomponent compounds provide richer compositional contrast and a broader descriptor manifold, which improves separability even when sample counts are small. Thus, TXL Fusion enhances robustness under chemical sparsity, but low-complexity topological materials remain intrinsically challenging for composition- and symmetry-level screening.

Prediction reliability was further assessed using top-label reliability diagrams and the Expected Calibration Error (ECE) (Supporting Information Section~S6.4). On the held-out test set, TXL Fusion attains higher accuracy than the standalone XGB model
(0.859 vs.\ 0.820) and markedly expands the high‑confidence prediction set: 6,229 of 7,397 test samples fall in the 0.9–1.0 confidence bin, compared to only 3,114 for XGB. This gain, however, is accompanied by an increase in overconfidence: the ECE grows from 0.023 (XGB) to 0.096 (TXL). In the top confidence bin, TXL exhibits a mean confidence of 0.992 and an accuracy of
0.911, whereas XGB achieves a mean confidence of 0.955 with an accuracy of 0.973. Hence, although TXL Fusion delivers superior accuracy and a substantially larger pool of high‑confidence predictions, its probabilities would benefit from post‑hoc recalibration to align confidence with empirical correctness.

\subsection{SHAP analysis}
\label{sec:shap}

To reveal the feature‑level origin of the TXL Fusion advantage, we conducted a
SHAP (SHapley Additive exPlanations) analysis of the weighted standalone XGB
and TXL Fusion models (Fig.~\ref{fig:feature_shap}). Global feature importance,
quantified by the mean absolute SHAP value, is shown in
Fig.~\ref{fig:feature_shap}A, C, while the beeswarm plots in
Fig.~\ref{fig:feature_shap}B, D show how each feature’s contribution varies
across individual samples, including its direction and spread.

For the XGB baseline, electron‑count parity (\texttt{Is\_total\_electrons\_even?},
mean $|\mathrm{SHAP}| = 0.700$) dominates, followed by SG priors
(\texttt{TI\_SG\_prob}: $0.266$, \texttt{SG}: $0.238$, \texttt{SM\_SG\_prob}:
$0.228$) and valence‑orbital descriptors (mean $p$: $0.196$, mean $d$: $0.148$).
The beeswarm plot reveals that even the most important features act
bidirectionally: depending on its value and interactions with other descriptors,
a single feature can push the prediction toward or away from a given class,
consistent with the factorized nature of the numerical descriptor space.

In TXL Fusion, the semantic embedding takes centre stage. The leading PCA
component of the SciBERT representation, \texttt{PCA‑0}, attains a mean
$|\mathrm{SHAP}|$ of $1.904$, far exceeding every other feature. It is followed
by the topogivity score \texttt{Trivial g(M)} ($0.569$), \texttt{PCA‑1}
($0.533$), electron‑count parity ($0.389$), and \texttt{PCA‑2} ($0.383$).
Semantic components such as \texttt{PCA‑12} ($0.139$) and physical descriptors
like \texttt{Trivial\_SG\_prob} ($0.119$) also contribute. The TXL beeswarm
plot shows that semantic coordinates generate the widest SHAP‑value ranges,
providing strong, sample‑dependent evidence. Heuristic scores, parity, and
symmetry priors remain present as interpretable anchors, but the model’s
decisions are now largely steered by the contextual semantic representation.

To interpret which handcrafted features favour each topological class, we
extracted the signed SHAP values from the XGB model and averaged them over
validation samples belonging to each true class
(Fig.~\ref{fig:xgb_class_support_shap}). For trivial compounds, the strongest
positive evidence comes from electron‑count parity ($0.328$), mean $p$‑valence
($0.235$), and the trivial SG prior ($0.209$), supplemented by
nonmetal and transition‑metal content. For TSMs, parity provides an
overwhelming positive signal ($0.916$), together with the semimetal SG prior
($0.240$), the SG index ($0.117$), and total electron count ($0.042$). For
TIs, the TI SG prior ($0.236$) and parity ($0.165$) are the leading
contributors, with weaker support from mean $p$ occupation, alkali‑metal
content, and the zero‑TI‑prior flag. These class‑conditioned patterns reveal a
relatively narrow evidentiary basis—mostly symmetry‑ and electron‑counting
signals—contrasting with the rich, context‑driven features supplied by the
semantic branch in TXL Fusion.

Overall, the SHAP analysis corroborates the design philosophy of TXL Fusion:
the semantic branch captures coupled physical information that is fragmented
across numerical descriptors, while the handcrafted features remain as
interpretable anchors. These feature‑level insights complement the branch‑level
ablation experiments: ablation quantifies the contribution of each modality,
whereas SHAP exposes the internal features that most strongly shape the fused
predictions.

\begin{figure}[htbp]
\centering
\includegraphics[width=1.0\linewidth]{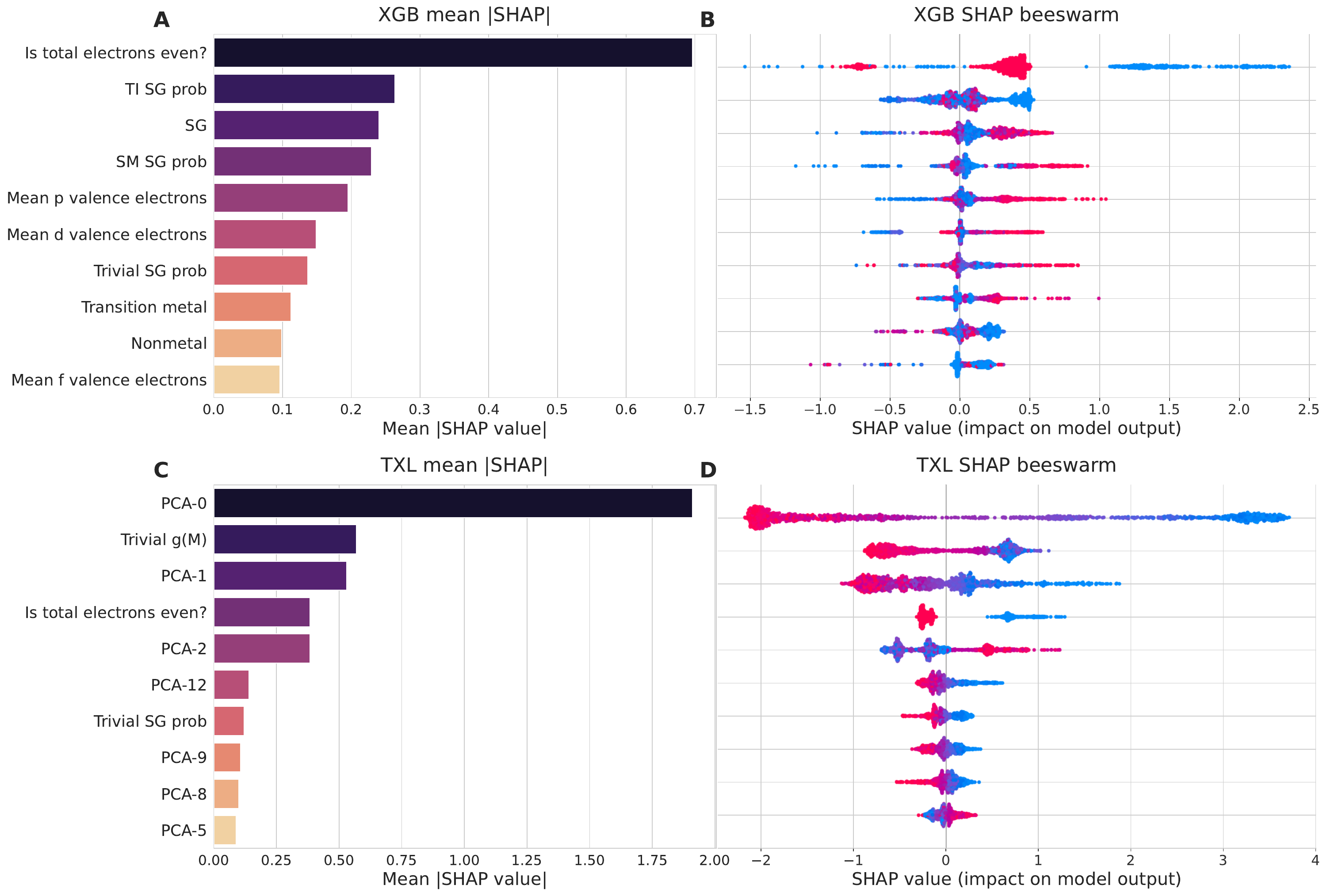}
\caption{\red{SHAP interpretation of the material classifiers on the validation set.
Panel A shows the mean absolute SHAP summary for the standalone XGB model, and
panel B shows the corresponding XGB beeswarm plot. Panel C shows the mean
absolute SHAP summary for TXL Fusion, and panel D shows the corresponding TXL
beeswarm plot. The mean-SHAP panels rank the most influential features globally,
whereas the beeswarm panels show the distribution and direction of SHAP
contributions across validation samples.}}
\label{fig:feature_shap}
\end{figure}

\begin{figure}
    \centering
    \includegraphics[width=0.77\linewidth]{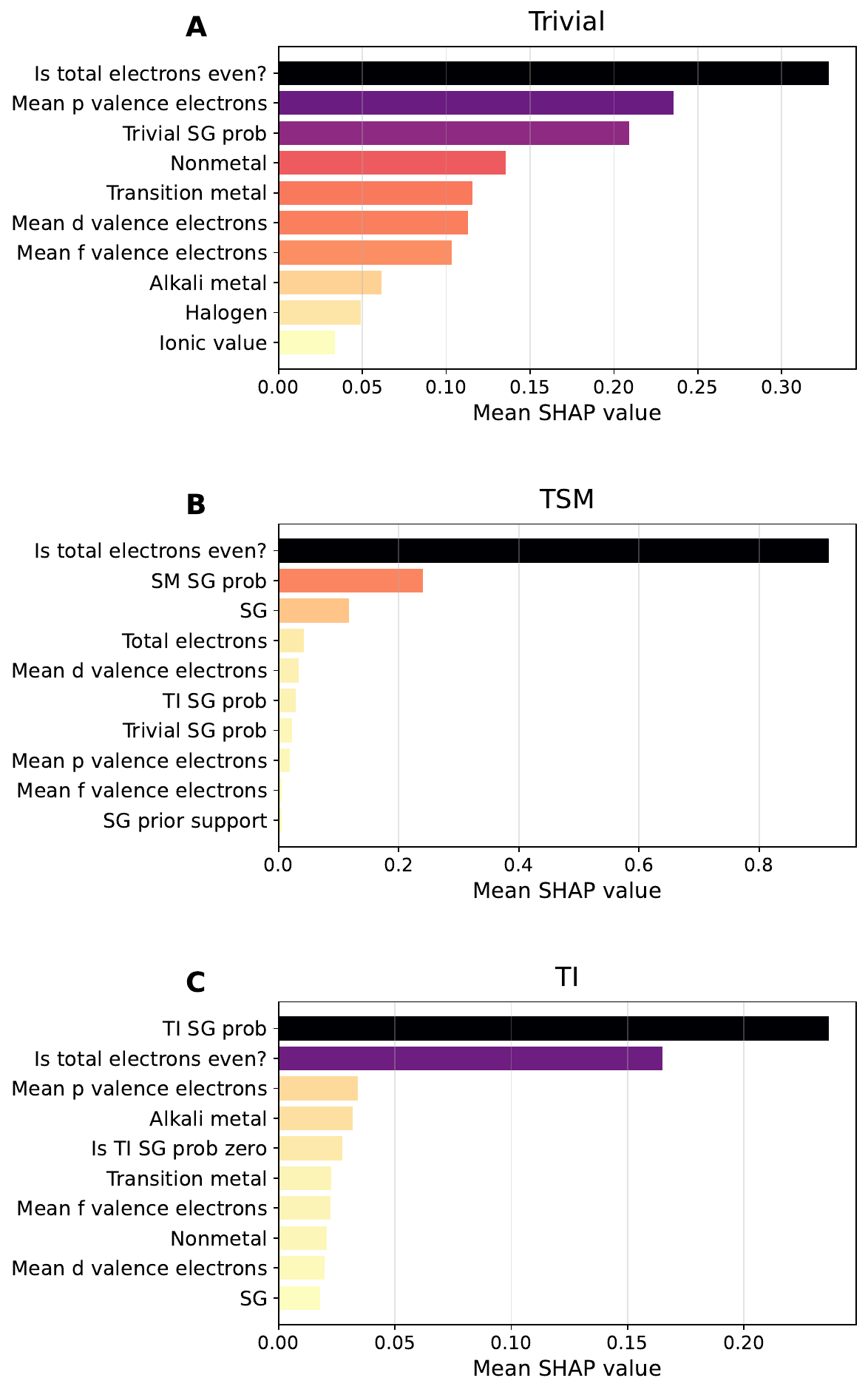}
    \caption{\red{Class-resolved positive SHAP evidence for the standalone XGB model on the validation set. Signed SHAP values were averaged over validation samples belonging to each true class, and the leading positive contributors are shown for (A) trivial materials, (B) TSMs, and (C) TIs.}}
    \label{fig:xgb_class_support_shap}
\end{figure}

\subsection{Semantic descriptor as the principal driver of TXL Fusion performance}
\label{sec:semantic_descriptor_driver}

To clarify why the semantic branch improves TXL Fusion, we compared the representation learned from fine-tuned SciBERT narratives with the corresponding numerical descriptor space (Supporting Information Section~S5). Conventional numerical descriptors encode materials as independent scalar features, including SG priors, electron counts, orbital occupations, bonding indicators, elemental ratios, and heuristic topogivity scores. Although physically interpretable, this factorized representation only indirectly captures coupled effects such as the joint dependence of topology on symmetry, orbital character, electron filling, and SOC-related chemistry. By contrast, the LLM semantic descriptor organizes the same evidence into a structured, physics-aware narrative before embedding, allowing these relationships to be encoded contextually rather than as isolated variables.

This difference is reflected in the geometry of the learned feature space. UMAP projections in Fig.~S5 (Supporting Information) show that the numerical descriptors provide only weak separation among trivial, TSM, and TI compounds, with silhouette scores of 0.0284 and 0.0257 using Euclidean and cosine metrics, respectively. In contrast, the SciBERT semantic embeddings compressed to 50 PCA components yield substantially clearer phase separation, with silhouette scores of 0.1246 and 0.1523 under the same metrics. Thus, the semantic descriptor acts as the main representational driver of TXL Fusion: it captures higher-order electronic--structural correlations that are difficult to express through handcrafted numerical descriptors alone, while the numerical and heuristic branches provide complementary interpretable physical constraints.

We further tested whether this semantic advantage requires the full 768-dimensional SciBERT embedding or can be retained in a compact representation (Supporting Information Section~S4). PCA fitted exclusively on the subtraining embeddings shows that the first 50 components retain 93.5\% of the total embedding variance, compared with 96.4\% for 100 components. Performance also saturates rapidly with dimensionality: the handcrafted-feature baseline without semantic embeddings gives validation macro-F1, weighted-F1, and accuracy values of 0.797, 0.847, and 0.845, whereas the 50-component representation reaches 0.821, 0.865, and 0.862, close to the full 768-dimensional embedding values of 0.826, 0.869, and 0.869. This compression decreases model complexity and training cost, mitigates unnecessary high-dimensional noise, and provides an efficient semantic descriptor for TXL Fusion without sacrificing most of the predictive benefit of the full embedding.

Feature-level interpretation further supports this conclusion. As shown in Fig.~\ref{fig:feature_shap}, the leading semantic component, \texttt{PCA-0}, has the largest mean absolute SHAP value, and additional semantic components also rank among the important features. At the same time, physically interpretable descriptors such as \texttt{Trivial g(M)}, electron-count parity, and \texttt{Trivial\_SG\_prob} remain influential. These results show that TXL Fusion does not replace chemical descriptors with a black-box semantic representation; rather, the semantic branch provides the dominant contextual signal, while heuristic and numerical descriptors supply complementary physically interpretable evidence.

\subsection{Statistical robustness of TXL Fusion}
\label{sec:statistical_robustness_main}

To test whether the improvement of TXL Fusion depends on a particular data partition, we repeated the complete training workflow across five independently generated stratified subtraining--validation splits while retaining the same cleaned held-out test set (Supporting Information Section~S8). TXL Fusion improved TI F1 in all five splits, with gains of 5.4, 3.8, 7.3, 6.3, and 3.8 percentage points over standalone XGB. Averaged across splits, TXL increased TI precision from $0.536 \pm 0.006$ to $0.626 \pm 0.023$, while maintaining comparable TI recall ($0.735 \pm 0.005$ for XGB versus $0.730 \pm 0.021$ for TXL), leading to an increase in TI F1 from $0.620 \pm 0.005$ to $0.673 \pm 0.013$. The robustness extends beyond the TI class: on the fixed held-out test set, TXL improves accuracy from $0.821 \pm 0.001$ to $0.852 \pm 0.006$, macro-F1 from $0.783 \pm 0.002$ to $0.816 \pm 0.007$, TSM F1 from $0.845 \pm 0.001$ to $0.869 \pm 0.006$, and trivial F1 from $0.885 \pm 0.000$ to $0.906 \pm 0.003$.

We further confirmed this advantage using paired-bootstrap resampling of the held-out test set across the five repeated splits (Supporting Information Section~S9). TXL produces a robust macro-F1 gain of $+0.0325$, with a strictly positive 95\% confidence interval of $[0.0240, 0.0410]$ and $P(\Delta>0)=1.0000$. The strongest class-specific gain is observed for TI precision ($+0.0898$, 95\% CI $[0.0719, 0.1078]$), whereas TI recall is statistically indistinguishable between TXL and XGB ($\Delta=-0.0057$, 95\% CI $[-0.0282, 0.0168]$). Thus, the TXL advantage is not an artifact of a favorable split or a small subset of held-out compounds; rather, semantic--numerical fusion consistently improves the reliability of TI assignments while also enhancing overall three-class classification performance.

\subsection{High-throughput screening of an external discovery space}
\label{sec:tsm_discovery}

To assess whether TXL Fusion can support candidate prioritization beyond the curated training and held-out test distributions, we applied the trained model to an external candidate space derived from Ma et al.\cite{ma2023topogivity}. This set was originally compiled by Tang et al.\cite{tang2019comprehensive} and contains materials whose topological character cannot be fully resolved from symmetry indicators alone. Among the 1,433 reported candidates, 1,235 compounds are already present in the topological-materials database, leaving 198 additional compounds outside our training database. Two compounds in SG~178 were excluded because this space group is absent from the cleaned topological-materials database used for training, resulting in a final external screening set of 196 candidates.

TXL Fusion was used as a high-throughput pre-screening tool to assign class probabilities to these 196 external candidates. The model predicted 24 materials as TSMs with varying confidence, while the remaining candidates were assigned to the trivial class; no materials were predicted as TIs. To focus downstream analysis on the most reliable candidates, we considered only high-confidence TSM predictions with $P_{\mathrm{TSM}}>0.90$. This threshold yielded 17 prioritized TSM candidates.

The high-confidence TSM candidates include LaMo$_2$O$_5$ (SG~186, $P_{\mathrm{TSM}}=0.999999$), Ta$_{21}$Te$_{13}$ (SG~183, $P_{\mathrm{TSM}}=0.999992$), Li$_{22}$Pb$_5$ (SG~196, $P_{\mathrm{TSM}}=0.998996$), In$_{11}$Mo$_{40}$O$_{62}$ (SG~26, $P_{\mathrm{TSM}}=0.998274$), Cl$_{11}$Mo$_3$N$_2$ (SG~29, $P_{\mathrm{TSM}}=0.995679$), Ge$_5$Li$_{22}$ (SG~196, $P_{\mathrm{TSM}}=0.990829$), CsC$_8$ (SG~180, $P_{\mathrm{TSM}}=0.990187$), Cs$_9$O$_3$Tl$_4$ (SG~197, $P_{\mathrm{TSM}}=0.988283$), Ag$_{10}$Br$_3$Te$_4$ (SG~36, $P_{\mathrm{TSM}}=0.987584$), Li$_{22}$Sn$_5$ (SG~196, $P_{\mathrm{TSM}}=0.968189$), Bi$_2$Cl$_7$Se$_5$ (SG~19, $P_{\mathrm{TSM}}=0.967146$), P$_3$Sc$_7$ (SG~186, $P_{\mathrm{TSM}}=0.953189$), AgPb$_4$Pd$_6$ (SG~152, $P_{\mathrm{TSM}}=0.942153$), InSr (SG~43, $P_{\mathrm{TSM}}=0.941447$), RbOS$_3$ (SG~150, $P_{\mathrm{TSM}}=0.914762$), P$_3$Rb$_2$Se$_6$ (SG~29, $P_{\mathrm{TSM}}=0.911584$), and KOS$_3$ (SG~150, $P_{\mathrm{TSM}}=0.910638$).

\begin{figure}
    \centering
    \includegraphics[width=1.0\linewidth]{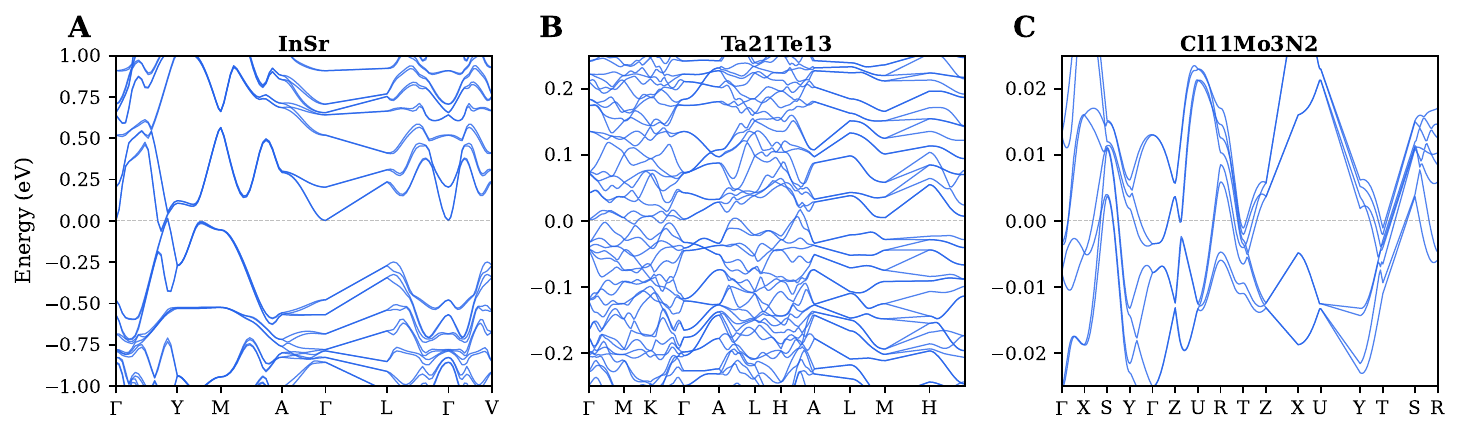}
    \caption{\red{Electronic band structures of selected TXL-prioritized candidates subjected to first-principles follow-up: (A) InSr (SG 43), (B) Ta$_{21}$Te$_{13}$ (SG 183), and (C) Cl$_{11}$Mo$_3$N$_2$ (SG 29).}}
    \label{fig:dft}
\end{figure}

Before first-principles follow-up, structural availability and database consistency were checked. In$_{11}$Mo$_{40}$O$_{62}$ and AgPb$_4$Pd$_6$ were assigned high TSM probabilities by TXL Fusion but could not be located in the Materials Project database,\cite{horton2025matproj} and were therefore not treated as directly actionable candidates. This step highlights an important limitation of descriptor-based screening: a model can identify a composition as topologically promising, but structural availability, database consistency, thermodynamic plausibility, and explicit electronic-structure validation remain necessary before a prediction can be regarded as a materials discovery.

For candidates with available and usable structural information, we performed DFT and Wannier-based follow-up valida tion for three selected TXL-prioritized materials, as shown in Fig.~\ref{fig:dft}. All three examined compounds are identified as Weyl semimetals. InSr (SG 43) hosts 43 pairs of Weyl nodes with chiralities of $\pm 1$ at general $k$-points in the Brillouin zone, with most Weyl nodes located within 0.1 eV of the Fermi level. Ta$_{21}$Te$_{13}$ (SG 183) exhibits a crystal-symmetry-protected twofold band crossing along the $\Gamma$--A high-symmetry line, involving bands 366 and 367. Cl$_{11}$Mo$_3$N$_2$ (SG 29) shows a pronounced near-Fermi-level Weyl-type band-contact feature along the $\Gamma$--X direction, where the valence and conduction bands form a weakly tilted crossing-like dispersion. In addition, InSr and Ta$_{21}$Te$_{13}$ have been independently validated as Weyl semimetals in Ref.~\citenum{ma2023topogivity}, providing external support for the TXL prioritization.

Overall, the external-screening analysis shows that TXL Fusion can prioritize plausible TSM candidates for more expensive DFT and Wannier-based validation. The workflow should therefore be viewed as a scalable triage strategy: it reduces the search space and identifies promising candidates, while retaining first-principles electronic-structure and topological analyses as the decisive validation step.
}

\section{Concluding remarks}
\red{
In this work, we introduced TXL Fusion, a hybrid semantic--numerical framework for the classification and discovery of topological materials. By integrating chemically informed heuristic scores, physically interpretable numerical descriptors, and LLM-derived semantic embeddings, TXL Fusion combines domain intuition with data-driven contextual representation. Across validation and held-out test evaluations, the model consistently outperforms the standalone numerical-descriptor XGB baseline and the direct heuristic $g(M)$ rule, demonstrating that no single information source is sufficient on its own. Instead, the strongest performance arises when compositional heuristics, symmetry- and electron-count descriptors, and semantic material narratives are fused within a unified learning framework.

A central result of this study is that the semantic branch acts as the main representational driver of TXL Fusion. The LLM-derived embeddings organize chemical formulae, SG information, orbital descriptors, bonding character, and heuristic reasoning into a context-aware representation that captures coupled electronic--structural relationships fragmented across conventional numerical features. This semantic information improves the discrimination of subtle topological regimes, while the numerical and heuristic branches retain interpretability and provide physically grounded constraints. The resulting framework improves overall accuracy, strengthens minority-class TI recognition, and provides a scalable pre-screening tool for external discovery spaces before expensive first-principles calculations or experimental validation.

Despite these advances, several intrinsic challenges remain. TIs are the most difficult class to predict, reflecting the combined effects of class imbalance, limited data support, subtle TI--TSM decision boundaries, and label ambiguity in the source database. As discussed in the Supporting Information, an internal consistency audit of the original 38,184-record dataset identifies 575 formula-plus-SG keys with conflicting class labels, 463 of which involve the TI class. For a composition-plus-SG learning task, such conflicts behave as contradictory supervision signals. They therefore provide a concrete source of label uncertainty, although they do not by themselves fully explain the remaining TI performance gap.

Element-resolved analysis further shows that low-complexity materials, especially unary and binary compounds, remain challenging. In these regimes, many compositional descriptors become constant or weakly discriminative, reducing descriptor separability precisely where subtle band-inversion physics becomes most important. Some low-complexity topological phases depend on symmetry-sensitive band ordering, orbital-projected band-edge character, strain effects, or wavefunction topology, none of which is explicitly encoded in composition-level descriptors. Future improvements will likely require incorporating more localized or symmetry-resolved electronic descriptors, such as symmetry-indicator eigenvalues, irreducible representations, Wannier charge centers, band-inversion indicators, or orbital-projected band-edge features. These quantities, however, generally require first-principles calculations and are not uniformly available at scale.

TXL Fusion should therefore be viewed as a high-throughput prioritization framework rather than a replacement for electronic-structure validation. Its strength lies in rapidly filtering large chemical spaces using inexpensive composition-, symmetry-, and text-derived information, thereby reducing the number of candidates requiring costly DFT calculations or experimental follow-up. Confidence and calibration analyses further indicate that the model produces a substantially expanded high-confidence decision regime, although post-hoc calibration may further improve probability reliability.

More broadly, this study shows that semantic--numerical fusion offers a practical route for materials discovery in regimes where physical knowledge is distributed across structured descriptors, empirical heuristics, and scientific language. Continued improvements in database fidelity, class balance, uncertainty calibration, and physics-aware representation learning should further enhance performance. Although developed here for topological materials, the TXL Fusion strategy is general and can be extended to other quantum and functional materials problems where interpretable physical descriptors and language-derived semantic knowledge provide complementary views of complex materials behavior.}

\begin{acknowledgement}

A.U. acknowledges funding from the National Natural Science Foundation of China (No. W2433037) and the Natural Science Foundation of Anhui Province (No. 2408085QA002). 

\end{acknowledgement}

\section{Code and Data Availability}

Code and data are available at \href{https://github.com/Arif-PhyChem/txl_fusion}{https://github.com/Arif-PhyChem/txl\_fusion}. \red{To facilitate broader community use, TXL Fusion is also planned to be deployed on the Aitomistic Hub in the near future (\href{https://aitomistic.xyz}{aitomistic.xyz}).}

\begin{suppinfo}
\red{Comprehensive methodology and supplementary analyses are provided in the Supporting Information. These include feature engineering and descriptor construction, TXL Fusion architecture and training details, semantic versus numerical descriptor comparisons, PCA-dimensionality sensitivity of SciBERT embeddings, element-resolved performance and calibration analyses, dedicated discussions of low-complexity compounds and chemical-sparsity generalization, class-imbalance sensitivity, repeated-split robustness and paired-bootstrap validation, DFT computational settings, and an assessment of noise and reliability in DFT-derived topological labels.}
\end{suppinfo}

\bibliography{references}

\end{document}


\maketitle

\vspace{2em}
\begin{center}
\renewcommand{\contentsname}{Table of Contents}
\tableofcontents
\end{center}
\vspace{1em}
\hrule
\vspace{1em}
\red{
\section{Features Analysis} \label{sec:data_analysis_si}
 
 \vspace{0.7cm}  

As described in the main text, the dataset used in this work was obtained from the topological materials database,\cite{topo_materials, bilbao_cryst, bradlyn2017topological, vergniory2019complete, vergniory2022all} which provides topological classifications based on electronic-structure calculations including spin--orbit coupling (SOC). The initial dataset contained 38,184 material records. Prior to analysis, descriptor construction, and model training, all chemical formulas were canonicalized, with deuterium ($\mathrm{D}$) consistently normalized to hydrogen ($\mathrm{H}$) throughout the workflow. This step ensures uniform elemental parsing, electron-count evaluation, and descriptor generation across all models.

To reduce label ambiguity, we removed formula--space group (SG) entries associated with contradictory class assignments (see Section~\ref{sec:labels-reliability} for more details). The resulting cleaned dataset contains 36,953 materials, including 5,587 topological insulators (TIs; $\sim$15.1\%), 13,652 topological semimetals (TSMs; $\sim$36.9\%), and 17,714 trivial materials ($\sim$47.9\%). This chemically and crystallographically diverse dataset spans a broad range of compositions and SGs, providing both a robust basis for analysing physically motivated descriptors and a stringent benchmark for developing machine-learning models for topological materials discovery. To ensure that the resulting models are accurate, comparable, and interpretable, we first examined which physically meaningful features most effectively distinguish TIs, TSMs, and trivial materials.

Guided by both theoretical considerations and systematic empirical analysis, we conducted a comprehensive feature selection process where our initial feature set spanned over many properties including chemical bonding characteristics (e.g., covalent vs. ionic tendencies), SOC strength ($\propto Z^4$), periodic table's group and column positions, total number of electrons, SG, valence electrons and atomic mass. Through iterative evaluation, we refined this broad feature pool to a compact set of descriptors that consistently offered both statistical robustness and physical interpretability.

Among these descriptors, SG symmetry emerged as one of the most decisive features, playing a central role in determining whether a compound is more likely to exhibit topological, semimetallic, or trivial electronic behavior. In the cleaned dataset used for this analysis, 216 unique SGs are represented. The class-resolved SG distributions reveal clear symmetry preferences, as shown in Fig.~\ref{fig:sg_dist}. Trivial compounds are most frequently found in SGs 14 (11.9\%), 62 (8.9\%), 2 (7.1\%), and 15 (6.5\%), indicating a strong bias toward lower-symmetry monoclinic/orthorhombic settings. In contrast, TSMs are concentrated in SGs 225 (8.0\%), 194 (7.8\%), 221 (7.7\%), and 139 (6.2\%), which are commonly associated with high-symmetry cubic or tetragonal structures. TIs most frequently occur in SGs 62 (11.4\%), 63 (8.0\%), 139 (7.9\%), and 12 (7.1\%), indicating a more nuanced balance between symmetry richness and topological permissiveness.

Crucially, Table~\ref{tab:zero_sg_prob} identifies SGs that are absent from one or more material classes in the cleaned dataset, reflecting symmetry settings that are strongly disfavored for particular electronic phases. For instance, SGs 196, 103, 106, 175, 210, and 211 are not associated with trivial compounds. Similarly, 33 SGs--including 3, 16, 17, 22, 24, 114, 145, 151, and 153--are not linked to semimetallic behavior. Most strikingly, 122 SGs do not host a single TI compound (e.g., 1--9, 16--46, 75--81, 90--92, 94--100, 102--110, 150--161, and 210--214; see Table~\ref{tab:zero_sg_prob} for the full list), highlighting the strong symmetry selectivity of topological insulating phases.

At the same time, SG symmetry is not an exclusive indicator of class membership. Fig.~\ref{fig:sg_dist} shows that several SGs--including 62, 63, 166, and 194--span multiple classes, indicating symmetry environments that can support different electronic behaviors depending on additional microscopic factors. For example, SGs such as 225, 227, 129, and 139 are frequently shared between trivial and TSM compounds but rarely host TI compounds, suggesting symmetry settings that often favor semimetallicity or conventional band structures. Conversely, SGs such as 2, 12, and 14 appear in both trivial and TI materials, showing that SG symmetry is highly informative but not sufficient by itself to determine topological character.

To complement symmetry, we broadened our analysis to include chemical and electronic descriptors beyond symmetry, as summarized in Table~\ref{tab:stat-analysis}. Among these, orbital occupancy patterns emerge as highly informative in distinguishing TIs from trivial and TSM classes. Notably, 31.9\% of TIs exhibit simultaneous d- and f-orbital occupancy, compared to 6.7\% in trivial compounds and 29.2\% in TSMs. This co-occupancy reflects enhanced SOC and complex orbital hybridization, conditions theoretically linked to band inversion and nontrivial topology. These observations are further supported by the average valence electron counts: TIs display substantially enhanced d-electron contributions (2.20) relative to trivial materials (0.79), though slightly lower than those of TSMs (2.49). For f-electrons, TIs exhibit a moderate increase over trivial compounds (0.80 vs.\ 0.13), while remaining marginally lower than TSMs (0.81), indicating that f-orbital involvement is important but not uniquely dominant in insulating topological phases. In contrast, p-orbital occupancy is reduced in TIs (1.32) and TSMs (1.20) compared to trivial materials (2.48), suggesting a departure from simple covalent bonding toward more correlated, relativistic electronic environments. Meanwhile, s-orbital contributions remain relatively consistent across all classes ($\sim$1.8), indicating that the distinctions in topological behavior are primarily driven by d-, f-, and p-orbital patterns.

In addition to orbital features, elemental composition provides key insights into the electronic character of materials. Both TIs and TSMs are enriched in transition metals (32.9\% and 36.9\%, respectively) and lanthanides (10.0\% and 10.5\%), elements with strong SOC and high-angular-momentum orbitals that facilitate band inversion and nontrivial topology. In contrast, trivial materials contain far fewer of these elements--11.6\% transition metals and 2.1\% lanthanides—consistent with weaker relativistic effects and simpler electronic structures.

The metalloid content also follows a similar pattern: TIs (16.2\%) and TSMs (11.4\%) exceed trivial compounds (8.2\%), reflecting their role in introducing intermediate bonding character and enhancing electronic complexity. Conversely, nonmetals dominate trivial materials (47.6\%) but are substantially less prevalent in TIs (20.5\%) and TSMs (18.6\%), suggesting that strongly covalent environments correlate with trivial phases, whereas topologically nontrivial systems favor heavier, more metallic elements with delocalized electrons and significant SOC.

Other elemental groups--including alkali metals, alkaline earth metals, and actinides--appear at low levels across all classes but are slightly more prevalent in nontrivial compounds. Halogens, general metals, and noble gases remain minor contributors, consistent with their limited involvement in stabilizing topological electronic structures.

Another feature, the total number of electrons per unit cell ($N_e$) plays a crucial factor in determining whether a material can exhibit a full band gap or must necessarily be metallic. In closed systems that preserve time-reversal symmetry, Kramers theorem ensures that all electronic bands are at least doubly degenerate.\cite{PhysRevB.105.L241106} As a result, systems with odd $N_e$ cannot achieve complete band filling and are thus compelled to be metallic or semimetallic. This theoretical expectation is borne out in our dataset: a substantial 72.0\% of TSMs feature odd $N_e$, reflecting their characteristic gapless nature and partially filled bands.

In contrast, this parity constraint on $N_e$ is less discriminative between trivials and TIs, both of which tend to have even electron counts that allow full band filling. Specifically, 96.0\% of trivial compounds and 85.5\% of TIs possess even $N_e$, indicating that electron count alone is insufficient to distinguish trivial from topological phases. 

In addition to already described features, we also include bonding characteristics, derived from the total electronegativity difference of constituent elements, as a meaningful compositional descriptor to characterize the chemical bonding nature of materials. This feature is categorized using established thresholds: compounds with an average difference $<$ 0.4 are labeled as very covalent, 0.4--1.0 as mostly covalent, 1.0--2.0 as moderately ionic, and $\ge$ 2.0 as highly ionic. This categorization captures the qualitative nature of electron sharing versus transfer, which can significantly affect the emergence of topological phases through orbital hybridization and gap formation. As shown in Table~\ref{tab:stat-analysis}, trivial materials tend to be more moderately ionic (58.9\%), whereas TIs and TSMs are more frequently mostly covalent (56.8\% and 55.1\%, respectively), suggesting that increased covalency--often tied to orbital delocalization and band inversion--plays a role in facilitating nontrivial topology. Very covalent bonding is also more prevalent in TIs (13.5\%) and TSMs (14.7\%)  compared to trivial ones (4.7\%), further supporting this trend. The low occurrence of highly ionic bonding across all categories (4.0\%, 1.2\%, and 1.7\% for trivials, TIs, and TSMs, respectively) implies that extreme ionicity may be generally unfavorable for topological features, possibly due to the localization of electronic states. 

Taken together, this analysis establishes a concise yet physically motivated set of features for characterizing topological materials, including orbital-resolved electron distributions, elemental composition trends, and SG symmetries, as summarized in Table~\ref{tab:stat-analysis}. With this foundation, we now proceed to develop and evaluate our approach that leverage these features for predictive classification of materials.

\begin{figure}
    \centering
    \includegraphics[width=1.0\linewidth]{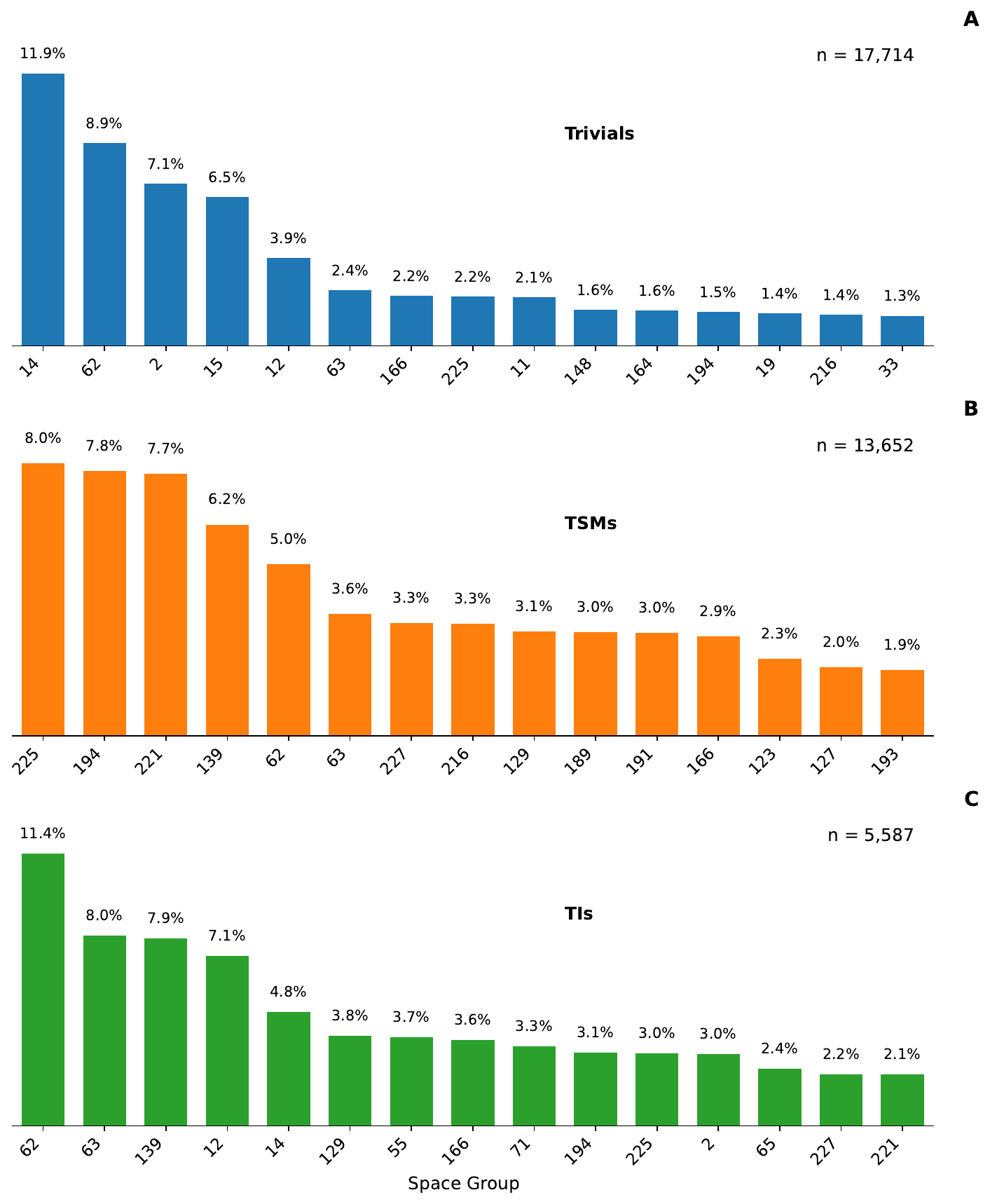}
    \caption{\red{SG distribution of (A) trivial, (B) TSM, and (C) TI compounds where $n$ represents the total number of compounds of a class.}}
    \label{fig:sg_dist}
\end{figure}

\begin{table}[h]
\caption{\red{SGs with zero probability in each material category.}}
\label{tab:zero_sg_prob}
\begin{adjustbox}{width=\textwidth}
\begin{tabular}{|p{2.5cm}|p{4.5cm}|p{7cm}|}
\hline
\textbf{Trivial} & \textbf{TSM} & \textbf{TI} \\
\hline
103, 106, 175, 196, 210, 211 & 3, 16, 17, 22, 24, 27, 32, 37, 39, 42, 45, 48--50, 77, 78, 80, 81, 94, 98, 105, 112, 114, 116, 145, 151, 153, 169, 177, 183, 192, 195, 208 & 1, 3--9, 16--46, 48, 49, 56, 68, 75--81, 90--92, 94--100, 102--110, 112, 114, 116, 118, 126, 132, 134, 143--146, 149--161, 169, 173--175, 177, 180--183, 185, 186, 188, 192, 195--199, 202, 203, 208, 210--214, 218, 219, 224\\
\hline
\end{tabular}
\end{adjustbox}
\end{table}

\begin{table}[htbp] 
\centering
\caption{\red{Comparative statistical analysis of material classes.}}
\label{tab:stat-analysis}
\begin{threeparttable}
\begin{tabular}{@{}>{\raggedright}p{3.5cm}lS[table-format=1.4]S[table-format=1.4]S[table-format=1.4]@{}}
\toprule
\textbf{Category} & \textbf{Subcategory} & \textbf{Trivial} & \textbf{TSM} & \textbf{TI} \\
\midrule

\multicolumn{5}{@{}l}{\textbf{Distribution (\%)}} \\
\cmidrule(l){2-5}
& Odd & 4.0 & {72.0} & {14.5} \\
& Even & 96.0 & {28.0} & {85.5} \\
\addlinespace

\multicolumn{5}{@{}l}{\textbf{Element Ratios (\%)}} \\
\cmidrule(l){2-5}
& Nonmetal & 47.6 & {18.6} & {20.5} \\
& Halogen & 12.0 & {3.3} & {2.1} \\
& Transition metal & 11.6 & {36.9} & {32.9} \\
& Alkali metal & 8.0 & {3.1} & {2.2} \\
& Metalloid & 8.2 & {11.4} & {16.2} \\
& Metal & 5.4 & {10.6} & {9.5} \\
& Alkaline earth metal & 4.5 & {3.6} & {4.8} \\
& Lanthanide & 2.1 & {10.5} & {10.0} \\
& Actinide & 0.5 & {1.8} & {1.9} \\
& Noble gas & 0.1 & {0.1} & {0.0} \\
\addlinespace
\multicolumn{5}{@{}l}{\textbf{Average valence electrons}} \\
\cmidrule(l){2-5}
& s & 1.82 & {1.80} & {1.82} \\
& p & 2.48 & {1.20} & {1.32} \\
& d & 0.79 & {2.49} & {2.20} \\
& f & 0.13 & {0.81} & {0.80} \\
\addlinespace
\multicolumn{5}{@{}l}{\textbf{d-f valence orbital statistics (\%)}} \\
\cmidrule(l){2-5}
& d \& f = 0 & 42.4 & {4.8} & {6.9} \\
& d $\neq$ 0 \& f = 0 & 48.8 & {58.1} & {53.3} \\
& d = 0 \& f $\neq$ 0 & 2.0 & {7.8} & {7.9} \\
& d \& f $\neq$ 0 & 6.7 & {29.2} & {31.9} \\
\addlinespace
\multicolumn{5}{@{}l}{\textbf{Bonding Characteristics (\%)}} \\
\cmidrule(l){2-5}
& Very covalent & 4.7 & 14.7 & 13.5 \\
& Mostly covalent & 31.3 & 55.1 & 56.8 \\
& Moderately ionic & 58.9 & 26.6 & 27.0 \\
& Highly ionic & 4.0 & 1.7 & 1.2 \\
\bottomrule
\end{tabular}

 \begin{tablenotes}
 \small
 \item Note: Percentages may not sum to 100\% due to rounding.
 \end{tablenotes}
\end{threeparttable}
\end{table}

\section{Details of the TXL Fusion framework}
\vspace{0.7cm}  
In this section, we provide a detailed description of each component of the TXL Fusion pipeline.

\subsection{Composition-based heuristic chemical rule module} \label{sec:tau}
\vspace{0.5cm}  
\begin{figure}
    \centering
    \includegraphics[width=1.0\linewidth]{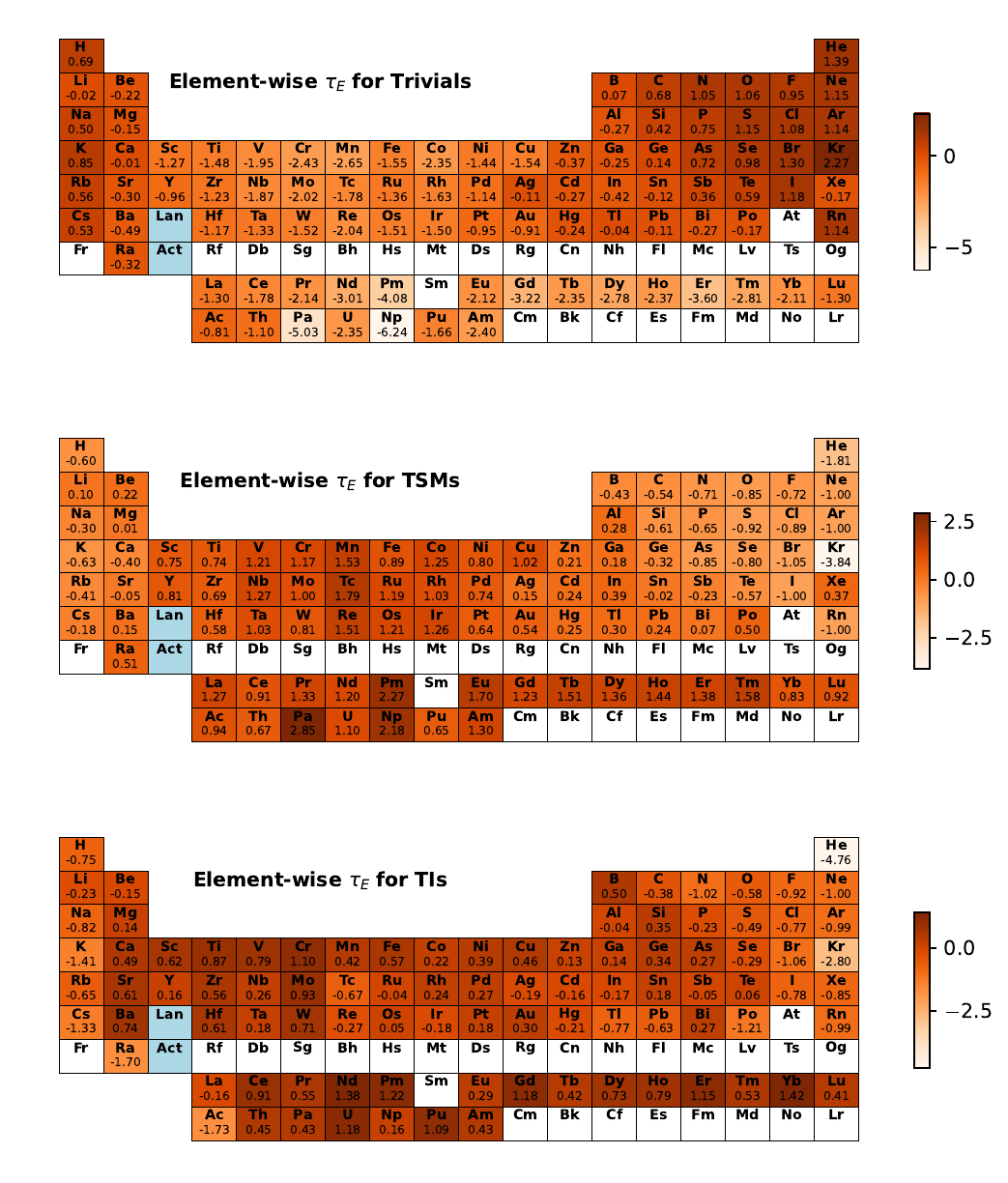}
    \caption{\red{Element-wise $\tau_E(c)$ maps for the class-balanced one-vs-rest composition heuristic: trivial-versus-rest, TSM-versus-rest, and TI-versus-rest.}}
    \label{fig:tau_pt}
\end{figure}

A framework introduced by Ma et al.~\cite{ma2023topogivity} proposes a composition-driven scoring system to assess the likelihood of a material exhibiting specific properties, such as topological behavior. While originally applied to topological materials classification, the approach is general and can be adapted to distinguish any two material categories. In this formulation, each chemical element is assigned a scalar value—referred to as its elemental contribution score (originally termed topogivity, $\tau_E$)—which quantifies its influence on the classification outcome. For a compound $M$, characterized by the relative abundance $f_E(M)$ of each element $E$, the overall material score $g(M)$ is computed as a weighted sum:

\begin{equation}
    g(M) = \sum_E f_E(M) \tau_E.
\end{equation}

A higher value of $g(M)$ suggests stronger membership in the positive class (e.g., TI), while a lower (typically negative) score indicates the opposite (e.g., trivial). The elemental scores $\tau_E$ are learned by training a linear classifier on a dataset of labeled materials. Each material is first represented as a compositional vector $\tilde{\mathbf{f}}(M)$, which lists the fractional contributions of all elements present (excluding one common element, such as oxygen, to avoid redundancy as it is the most abundant element and can be retrieved using the normalization constraint $\sum_E f_E(M) = 1$). The elements are ordered by increasing atomic number to ensure consistency across the dataset:

\begin{equation} \label{eq:tau_vector}
    \tilde{\mathbf{f}}(M) = \left(f_{\text{Li}}(M), f_{\text{Be}}(M), \ldots, f_{\text{U}}(M)\right).
\end{equation}

In our implementation, this results in a 91-dimensional input vector (with 92 different chemical elements in our dataset). The classification task is framed as a binary problem, where one class is labeled as $+1$, and other classed as $-1$. A soft-margin linear support vector machine (SVM) is then trained to discriminate between the two categories. The decision function takes the form:

\begin{equation}
    Q(M) = \mathbf{w}^T \tilde{\mathbf{f}}(M) + b,
\end{equation}

where $\mathbf{w}$ is the weight vector encoding learned elemental contributions and $b$ is a scalar bias. The model is optimized by minimizing a regularized hinge loss:
\begin{equation}
    \min_{\mathbf{w}, b} \left[ \frac{1}{N} \sum_{i=1}^{N} \max(0, 1 - y^{(i)} Q(M^{(i)})) + \gamma \|\mathbf{w}\|^2 \right],
\end{equation}

where $\gamma$ is a regularization parameter, $y^{(i)} \in {+1, -1}$ denotes the class label of material $M^{(i)}$, and $N$ is the total number of training examples. After training, the elemental score for each element is extracted from the learned weights as:
\red{
\begin{equation} \label{eq:tau_mapping}
\tau_E = \begin{cases}
b, & \text{if } E = \tilde{E} \text{ (excluded element)} \\
w_{\iota(E)} + b, & \text{otherwise},
\end{cases}
\end{equation}
where $\iota(E)$ maps each element to its corresponding index in the compositional vector $\tilde{\mathbf{f}}(M)$. This simple yet interpretable model captures general chemical trends and provides a compact feature for materials classification. In our study, we adapt the binary $g(M)$ rule to the present three-class problem by training three class-balanced one-vs-rest linear SVMs: Trivial-versus-rest, TSM-versus-rest, and TI-versus-rest. Each model assigns an interpretable elemental weight $\tau_E(c)$ for class $c \in \{\mathrm{Trivial},\mathrm{TSM},\mathrm{TI}\}$. For a compound $M$, these maps define class-specific composition scores,
\begin{equation}
   g_c(M) = \sum_E f_E(M)\,\tau_E(c), \qquad c \in \{\mathrm{Trivial},\mathrm{TSM},\mathrm{TI}\},
\end{equation}
which can be collected into a three-dimensional score vector
\begin{equation}
   \boldsymbol{g}(M) = [g_{\mathrm{Trivial}}(M),\; g_{\mathrm{TSM}}(M),\; g_{\mathrm{TI}}(M)] .
\end{equation}
The direct three-class extension assigns the final heuristic label by selecting the largest of the three scores,
\begin{equation}
   \hat{y}(M) = \arg\max_{c \in \{\mathrm{Trivial},\mathrm{TSM},\mathrm{TI}\}} g_c(M).
\end{equation}
Thus, the three-class rule remains a purely composition-based and interpretable extension of the original topogivity framework; no secondary classifier, probability calibration, or threshold tuning is introduced.

Fig.~\ref{fig:tau_pt} presents periodic table distributions of $\tau_E(c)$ for all three one-vs-rest maps, namely $g_{\mathrm{Trivial}}(M)$, $g_{\mathrm{TSM}}(M)$, and $g_{\mathrm{TI}}(M)$. The learned weights retain clear chemical interpretability. In the $g_{\mathrm{Trivial}}(M)$ map, light nonmetals, halogens, and other elements frequently associated with chemically simple trivial compounds receive high positive scores. In the $g_{\mathrm{TSM}}(M)$ map, heavier elements and transition-metal/semimetal
chemistry receive stronger positive contributions, consistent with the prevalence
of strong SOC coupling and partially filled bands in TSMs. The $\tau_E(c)$ values of
$g_{\mathrm{TI}}(M)$ map is also shown for completeness, as it enters the
three‑score $\arg\max$ reference heuristic; it exhibits a pattern similar to the
TSM map, with heavy and transition‑metal elements likewise receiving positive
weights. Despite this resemblance, composition‑based TI classification remains
weak: the small number of TI training examples and the lack of a clear
composition‑only fingerprint that reliably separates TIs from trivial and TSM
compounds limit the selectivity of this channel.}

\subsection{Numerical descriptor module} \label{sec:num_desc}
\vspace{0.7cm}  

The numerical descriptor module provides a transparent and physically
interpretable representation of each material by combining compositional,
electronic, bonding, and crystallographic information into a fixed-length
feature vector. Unlike purely heuristic composition-based rules, this framework
encodes each compound using quantitative descriptors derived from elemental
properties and SG statistics. These descriptors are then used as input
features for the numerical machine-learning classifier.

For a material $M$, we define the numerical descriptor vector as
\begin{equation}
    \mathbf{x}(M) \in \mathbb{R}^{d},
\end{equation}
where each component corresponds to a chemically or structurally meaningful
feature. The SG number is included directly as
\begin{equation}
    x_{\mathrm{SG}}(M) = \mathrm{SG}(M),
    \qquad
    \mathrm{SG}(M) \in \{1,\ldots,230\}.
\end{equation}

The total number of electrons is computed from the elemental composition:
\begin{equation}
    N_e(M) =
    \sum_{i=1}^{n_M} N_i Z_i,
\end{equation}
where $N_i$ is the number of atoms of element $E_i$, $Z_i$ is its atomic
number, and $n_M$ is the number of distinct elements in the compound. To
capture parity-related effects, we further define
\begin{equation}
\delta_{\mathrm{even}}(M)=
\begin{cases}
1, & N_e(M) \bmod 2 = 0,\\
0, & \mathrm{otherwise}.
\end{cases}
\end{equation}
This feature is motivated by the relationship between electron filling,
band degeneracy, and possible semimetallic behavior in crystalline solids.

Orbital-resolved valence information is represented by the average number of
valence electrons in the $s$, $p$, $d$, and $f$ channels:
\begin{equation}
    \mu_o(M)
    =
    \frac{1}{N_{\mathrm{at}}}
    \sum_{i=1}^{n_M} N_i \nu_o(E_i),
    \qquad
    o \in \{s,p,d,f\},
    \label{eq:mean_orbital_valence}
\end{equation}
where $\nu_o(E_i)$ denotes the number of valence electrons of element $E_i$
in orbital channel $o$, and
\begin{equation}
    N_{\mathrm{at}} = \sum_{i=1}^{n_M} N_i
\end{equation}
is the total number of atoms in the formula. For transition-metal elements,
$(n-1)d$ electrons are included as valence electrons, while for lanthanides
and actinides the corresponding $f$-electron contributions are also included.

From these orbital descriptors, we define binary indicators for the presence
of $d$ and $f$ valence electrons:
\begin{equation}
\begin{aligned}
    \delta_d(M)  &= \mathbb{I}\!\left[\mu_d(M)>0\right],\\
    \delta_f(M)  &= \mathbb{I}\!\left[\mu_f(M)>0\right],\\
    \delta_{df}(M) &= \mathbb{I}\!\left[
        \delta_d(M)=1 \ \mathrm{and}\ \delta_f(M)=1
    \right].
\end{aligned}
\label{eq:df_flags}
\end{equation}
These descriptors distinguish materials containing transition-metal,
lanthanide, or actinide-derived electronic states, which are often relevant
to band inversion, narrow-band physics, and spin-orbit-coupled electronic
structures.

Bonding character is described using the average pairwise electronegativity
difference among the constituent elements:
\begin{equation}
    \Delta \chi(M)
    =
    \frac{2}{n_M(n_M-1)}
    \sum_{i<j}
    \left|
    \chi(E_i)-\chi(E_j)
    \right|.
    \label{eq:ionic_value}
\end{equation}
For single-element compounds, or cases where the bonding character is not
well defined, this descriptor is treated as undefined. The continuous
electronegativity-difference descriptor is supplemented by a one-hot bonding
category vector,
\begin{equation}
\begin{split}
    \mathbf{b}(M) =
    [&b_{\mathrm{undef}},
    b_{\mathrm{very\ covalent}},
    b_{\mathrm{mostly\ covalent}},\\
    &b_{\mathrm{moderately\ ionic}},
    b_{\mathrm{highly\ ionic}} ].
\end{split}
\label{eq:bonding_vector}
\end{equation}
The bonding categories are assigned according to the following thresholds:
very covalent for $\Delta\chi < 0.4$, mostly covalent for
$0.4 \leq \Delta\chi < 1.0$, moderately ionic for
$1.0 \leq \Delta\chi < 2.0$, and highly ionic for
$\Delta\chi \geq 2.0$.

To incorporate crystallographic priors, we compute class-conditional
SG statistics from the subtraining data. For a SG $g$, the
empirical probability of observing class
$c \in \{\mathrm{trivial},\mathrm{TSM},\mathrm{TI}\}$ is defined as
\begin{equation}
    P_c^{(g)}
    \equiv
    P(y=c \mid \mathrm{SG}=g)
    =
    \frac{N_c^{(g)}}{N^{(g)}},
    \label{eq:sg_prior}
\end{equation}
where $N_c^{(g)}$ is the number of subtraining materials in SG $g$
with class label $c$, and $N^{(g)}$ is the total number of subtraining
materials observed in that SG. The resulting SG-prior features are
\begin{equation}
    \mathbf{p}_{\mathrm{SG}}(M)
    =
    \left[
    P_{\mathrm{trivial}}^{(g)},
    P_{\mathrm{TSM}}^{(g)},
    P_{\mathrm{TI}}^{(g)}
    \right],
    \qquad g=\mathrm{SG}(M).
\end{equation}
We also include the support count of the SG prior,
\begin{equation}
    S_{\mathrm{SG}}(g)=N^{(g)},
\end{equation}
which indicates how many subtraining samples contributed to the corresponding
SG statistics. In addition, zero-probability indicators are included:
\begin{equation}
    z_c^{(g)}
    =
    \mathbb{I}
    \left[
    P(y=c \mid \mathrm{SG}=g)=0
    \right],
    \qquad
    c \in \{\mathrm{trivial},\mathrm{TSM},\mathrm{TI}\}.
    \label{eq:sg_zero_flags}
\end{equation}
These features allow the model to exploit symmetry-dependent class tendencies
while retaining information about the statistical reliability of each
SG prior.

To encode broad chemical-family information, each element is assigned to one
of several periodic-table categories:
\begin{equation}
\begin{split}
\mathcal{C} =
\{&
\mathrm{alkali},\
\mathrm{alkaline\ earth},\
\mathrm{nonmetal},\
\mathrm{halogen},\
\mathrm{noble\ gas},\\
&\mathrm{transition\ metal},\
\mathrm{metal},\
\mathrm{lanthanide},\
\mathrm{actinide},\
\mathrm{metalloid}
\}.
\end{split}
\label{eq:element_categories}
\end{equation}
For each category $c \in \mathcal{C}$, we compute the atomic percentage
\begin{equation}
    \phi_c(M)
    =
    100
    \times
    \frac{1}{N_{\mathrm{at}}}
    \sum_{i=1}^{n_M}
    N_i \mathbb{I}[E_i \in c].
    \label{eq:category_fraction}
\end{equation}
The vector $(\phi_c)_{c\in\mathcal{C}}$ summarizes the distribution of
chemical families within the compound and provides a compact representation
of broad compositional trends.

The final descriptor vector is obtained by concatenating all feature groups:
\begin{equation}
\begin{aligned}
\mathbf{x}(M) =
\big[
&\mathbf{x}_{\mathrm{basic}},
\mathbf{x}_{\mathrm{orb}},
\mathbf{x}_{\mathrm{bond}},
\mathbf{x}_{\mathrm{SG}},
\mathbf{x}_{\mathrm{cat}}
\big],
\end{aligned}
\label{eq:descriptor_groups}
\end{equation}
where the individual blocks are defined as
\begin{align}
\mathbf{x}_{\mathrm{basic}}
&=
\big[
\mathrm{SG},
N_e,
\delta_{\mathrm{even}}
\big],
\\[2mm]
\mathbf{x}_{\mathrm{orb}}
&=
\big[
\mu_s,
\mu_p,
\mu_d,
\mu_f,
\delta_d,
\delta_f,
\delta_{df}
\big],
\\[2mm]
\mathbf{x}_{\mathrm{bond}}
&=
\big[
\Delta\chi,
\mathbf{b}
\big],
\\[2mm]
\mathbf{x}_{\mathrm{SG}}
&=
\big[
P_{\mathrm{trivial}}^{(g)},
P_{\mathrm{TSM}}^{(g)},
P_{\mathrm{TI}}^{(g)},
S_{\mathrm{SG}},
z_{\mathrm{trivial}}^{(g)},
z_{\mathrm{TSM}}^{(g)},
z_{\mathrm{TI}}^{(g)}
\big],
\\[2mm]
\mathbf{x}_{\mathrm{cat}}
&=
\big[
\phi_c
\big]_{c\in\mathcal{C}}.
\end{align}
This numerical representation combines electron-count descriptors,
orbital-resolved valence information, bonding characteristics,
SG-derived statistical priors, and chemically interpretable
compositional summaries. It therefore provides a physically motivated and
transparent input representation for the numerical machine-learning
classifier.

\subsection{LLM: Semantic descriptor based module}
\vspace{0.7cm}  

Large language models (LLMs) unify structured data with domain knowledge into a cohesive, high-dimensional representation, where self-attention captures coupled physical effects—such as the interplay between crystalline symmetry and spin–orbit coupling—as context-aware vectors, in contrast to conventional numerical descriptors that factorise material properties into isolated scalars. By leveraging SciBERT~\cite{beltagy2019scibert} pretrained on an extensive scientific corpus, the embeddings inherit latent physical regularities, enabling the classification pipeline to exploit complex nonlinear correlations between orbital character and band topology that are often inaccessible through hand-crafted features (a comparative discussion of semantic and numerical descriptors is provided in Section~\ref{sec:sem-vs-num}).

The semantic descriptor module translates the numerical and heuristic descriptors defined in Sections~\ref{sec:tau} and~\ref{sec:num_desc} into a structured scientific narrative. For each compound, the chemical formula is canonicalised with \texttt{pymatgen} using the standard deuterium-to-hydrogen normalisation, and target labels are mapped to three unified classes: trivial, TSM, and TI. Each material is serialised using a fixed template containing the normalised formula, SG number, SG class priors, zero-probability class indicators, element-category percentages, mean orbital-resolved valence occupations, orbital-presence flags, total electron count and parity, bonding character, and composition-based topogivity scores. This template preserves the interpretability of the numerical descriptors while allowing the transformer model to learn contextual relationships: SG fields encode symmetry-dependent class tendencies, element categories summarise broad chemical-family composition, $d/f$-orbital flags capture transition-metal, lanthanide, and actinide electronic character, electron-count parity and bonding supply physically motivated cues, and topogivity scores introduce transparent composition-derived heuristics for the trivial and TSM classes.

The final semantic representation can be written compactly as
\begin{equation}
\begin{aligned}
\mathcal{T}(M)=
\{&
\mathrm{formula},
\mathrm{SG},
\mathbf{p}_{\mathrm{SG}},
\mathbf{z}_{\mathrm{SG}},
\boldsymbol{\phi},
\boldsymbol{\mu}_{spdf},
\boldsymbol{\delta}_{df},\\
&N_e,
\delta_{\mathrm{even}},
\mathrm{bonding},
\mathbf{g},
\mathrm{score\ patterns}
\},
\end{aligned}
\label{eq:semantic_template}
\end{equation}
where $\mathcal{T}(M)$ denotes the structured text representation of material
$M$. The quantities $\mathbf{p}_{\mathrm{SG}}$,
$\mathbf{z}_{\mathrm{SG}}$, $\boldsymbol{\phi}$,
$\boldsymbol{\mu}_{spdf}$, $\boldsymbol{\delta}_{df}$, $N_e$, and
$\delta_{\mathrm{even}}$ follow the definitions in
Section~\ref{sec:num_desc}. The vector $\mathbf{g}$ contains the
composition-based topogivity scores, and the score-pattern terms encode the
sign and relative relationship between the trivial and TSM topogivity scores
(the TI topogivity score was excluded due to its limited selectivity in the
composition-only analysis).

The generated semantic records were stored as JSON entries containing the
normalised compound formula, SG number, structured text sequence, and
target label. To guarantee consistency with the numerical-descriptor workflow,
all semantic records were validated against the shared split manifest,
confirming identical compound identity, SG number, class label, and
row ordering across the two descriptor modalities.

We used \texttt{scibert\_scivocab\_uncased}\cite{beltagy2019scibert} as the
language-model backbone because it is pretrained on scientific literature and
employs a domain-specific vocabulary. The structured semantic narratives were
used for fine-tuning, enabling the model to learn from physically organised
textual descriptors rather than unconstrained free text. Training details are
provided in the following section.

\section{Training details}
\vspace{0.7cm}  

This section describes the training procedures for all models employed in this
study, including the composition-based heuristic model, the standalone XGBoost
(XGB) classifier, and the TXL Fusion framework. As described in Section~\ref{sec:data_analysis_si}, the original dataset contained
38,184 material records, comprising
18,090 ($\sim$47.3\%) trivial materials,
13,985 ($\sim$36.6\%) TSMs, and
6,109 ($\sim$16\%) TIs. Before feature
construction and model training, all chemical formulas were canonicalized, with
deuterium ($\mathrm{D}$) normalized to hydrogen ($\mathrm{H}$) throughout
the entire workflow. This normalization was applied consistently to the
composition-based heuristic model, numerical descriptor generation, semantic
descriptor construction, and all downstream training, validation, and testing
procedures.

To reduce label inconsistency, we removed formula--SG entries assigned
to contradictory topological labels (see Section~\ref{sec:labels-reliability} for more details). After this cleaning step, the dataset
contained 36,953 materials, including
17,714 trivial materials ($\sim$47.9\%),
13,652 TSMs ($\sim$36.9\%), and
5,587 TIs ($\sim$15.1\%). Of these, 29,556
records (80.0\%) formed the cleaned training pool, while 7,397 records (20.0\%) were reserved as
an independent held-out test set. The cleaned training pool was further divided
using a shared stratified split into 23,644 subtraining records (80.0\% of the training pool; 64.0\% of the cleaned dataset) and 5,912
validation records (20.0\% of the training pool; 16.0\% of the cleaned dataset). The same validation and held-out test protocol was used for
all model comparisons to ensure a consistent and unbiased assessment. The
following subsections describe the feature preparation, model configuration,
hyperparameter selection, and training strategy for each model.

\subsection{Composition-based heuristic chemical rule} \label{sec:tau_training}
\vspace{0.7cm}  

For the composition-based $g(M)$ heuristic, each compound is represented solely by its chemical formula and then converted into the non-oxygen elemental-fraction vector $\tilde{\mathbf{f}}(M)$ (Eq.~\eqref{eq:tau_vector}). To extend the original binary topogivity formulation to the present three-class task, we train three one-vs-rest linear SVMs: trivial-versus-rest, TSM-versus-rest, and TI-versus-rest. In each binary fit, the target class is assigned $+1$ and the pooled remaining two classes are assigned $-1$. All three SVMs are trained on the same 23,644-record subtraining split, and for a given grid value of $\gamma$ the soft-margin parameter is set to $C = 1/(N_{\text{subtrain}}\gamma)$, where $N_{\text{subtrain}}$ is the total number of subtraining samples (23,644) entering that binary one-vs-rest fit.

\red{Class imbalance is handled separately by balanced class weights in the hinge-loss objective. Specifically, for each binary SVM, scikit-learn assigns class weights:
\begin{equation} \label{eq:weight_balance}
    w_c = \frac{N_{\text{subtrain}}}{Kn_c},
\end{equation}
 where $K=2$ is the number of classes in the binary problem, and $n_c$ the
number of samples in class $c$; consequently, the minority positive class
(most notably TI in the TI‑versus‑rest fit) receives a larger loss weight than
the pooled majority class (a comprehensive
sensitivity analysis of this and alternative weighting strategies—including
additional TI‑specific overweighting—and their effects on both global and
per‑class performance is provided in Section~\ref{sec:class-imbalance}.)}. The value of $\gamma$ is selected on the shared validation split by maximizing the Macro-F1 of the complete three-score heuristic, with validation TI F1 and then accuracy used only as tie-breakers. This procedure selected $\gamma = 1.28\times10^{-8}$, corresponding to $C = 3304.22$ for $N=23,644$. The selected SVM coefficients define the elemental weights $\tau_E(c)$ and the class-specific topogivity scores $g_c(M)$ while the direct heuristic prediction is simply $\arg\max_c [g_c(M)]$. 

\subsection{Standalone XGB model} \label{sec:training_xgb_model}
\vspace{0.7cm}  

A standalone XGBoost (XGB) classifier was trained on the numerical descriptors
detailed in Section~\ref{sec:num_desc}. All chemical formulas were canonicalized
with the same deuterium ($\mathrm{D}$) normalized to hydrogen ($\mathrm{H}$) and
aligned to the shared split manifest used throughout this study: 23,644 training
and 5,912 validation samples. SG class priors were calculated
exclusively from the training partition and applied to both sets, preventing
leakage while preserving symmetry-dependent statistical information.

The feature vector for each material comprised the SG number, total
electron count, electron-count parity, mean $s$, $p$, $d$, and $f$
valence occupations, binary $d$- and $f$-orbital presence flags, bonding
descriptors derived from electronegativity differences, empirical SG
class priors, zero-probability SG indicators, SG support
counts, and element-category percentages. The complete feature table was saved
before training to guarantee full reproducibility.

\red{Labels were encoded with a \texttt{LabelEncoder} over the three classes
\{\texttt{trivial}, \texttt{TSM}, \texttt{TI}\}. To mitigate
class imbalance, each sample was weighted by the inverse class frequency in the
training set (Eq.~\eqref{eq:weight_balance}) with $K = 3$ (see
Section~\ref{sec:class-imbalance} for a sensitivity analysis on class imbalance). This weighting scheme yielded weights of 0.898 for TSMs, 2.238 for TIs, and 0.694 for trivial, increasing the contribution of the minority topological class during model fitting. These weights were passed directly
to the XGB training procedure.}

The XGB model was configured with a maximum depth of 4, a learning rate of 0.01,
1,000 estimators, subsample ratio of 0.7, column subsample ratio of 0.7, minimum
child weight of 3, split-gain threshold $\gamma = 0.2$, L1 regularization
$\alpha = 0.2$, and L2 regularization $\lambda = 2.0$. Multi-class log loss
was used for validation, with early stopping after 20 rounds without improvement.
Reproducibility was enforced by setting \texttt{random\_state = 42}. Training
employed the histogram tree method with GPU acceleration; if unavailable, an
identical CPU configuration was automatically invoked.

After training, the booster was moved to the CPU for validation to avoid
device-mismatch warnings. Performance was assessed via accuracy, macro-averaged
F1 score, weighted F1 score, per-class precision, recall, F1 score, and the
confusion matrix. The trained model, label mapping, feature names, class
weights, validation metrics, SG priors, and feature importance values
were saved to disk. The final booster was stored in JSON format, and feature
importances were exported as structured data and as a PDF visualization for
downstream analysis.

\subsection{Finetuning of LLM for TXL Fusion} \label{sec:finetuning_llm}
\vspace{0.7cm}  

The semantic branch of TXL Fusion builds on
\texttt{allenai/scibert\_scivocab\_uncased}, a BERT-style encoder pretrained
on scientific text (12 attention heads, hidden size 768). A three‑class
sequence‑classification head was added and fine‑tuned on the curated narrative
corpus generated from the training pool (trivials, TSMs, TIs). The
independent test set was entirely withheld from fine‑tuning, checkpoint
selection, and early stopping.

All narratives were aligned to the shared stratified split manifest by
normalised formula, SG number, and label, ensuring exact parity with
the numerical and heuristic branches. Only the structured text was tokenised
(SciBERT tokenizer, max length 512 tokens); the label served solely as the
supervision target and was not prepended to the input. Explicit label$\leftrightarrow$index
mappings kept the output interpretation consistent.

To counteract class imbalance, we minimised a weighted cross‑entropy loss whose
class weights were computed exclusively from the training split
(Eq.~\eqref{eq:weight_balance}).

Fine‑tuning used the Hugging Face \texttt{Trainer} (seed 42): up to 8 epochs,
learning rate $2\times10^{-5}$, weight decay 0.01, batch size 16 per device.
The best checkpoint was selected by the highest validation weighted‑F1, with
early stopping patience of 5 evaluation epochs.

\subsection{Training of TXL Fusion} \label{sec:txl_fusion_training}
\vspace{0.7cm}  

TXL Fusion is a late‑fusion classifier that integrates three complementary feature blocks: composition‑based heuristics, numerical descriptors, and semantic embeddings from the fine‑tuned SciBERT model.

The numerical/heuristic block re-uses a curated subset of the descriptors
defined in Sections~\ref{sec:num_desc} and~\ref{sec:tau}. This 25-dimensional
block includes the selected topogivity scores
$g_{\mathrm{trivial}}(M)$ and $g_{\mathrm{TSM}}(M)$ and their sign-pattern
indicators, together with SG number and SG priors, SG-prior support,
electron-count and parity descriptors, selected orbital descriptors, bonding
indicators, and selected element-category percentages. As noted earlier, the
TI topogivity score was excluded owing to its limited selectivity. The semantic
block employs the encoder described in Section~\ref{sec:finetuning_llm}: each
structured narrative is encoded and mean-pooled to a 768-dimensional embedding,
then compressed by PCA, fitted exclusively on the subtraining split, to the
first 50 principal components (see Section~\ref{sec:pca_variance}). 

Fusion proceeds through a learned gating mechanism followed by hierarchical
probabilistic refinement. The semantic and numerical streams are independently
projected into a 128-dimensional hidden space with ReLU activation and dropout
of 0.2, and are combined by sample-dependent gating weights before three-class
prediction. This gated module is trained with AdamW using a learning rate of
$10^{-3}$, weight decay $10^{-4}$, batch size 256, and a maximum of 80 epochs.
Early stopping is based on validation macro-F1 with a patience of 10 epochs.

A two-stage hierarchical XGBoost classifier is then applied to the fused
representation. The first stage separates trivial from nontrivial materials,
and the second stage, trained only on the nontrivial subset, discriminates
TSM from TI. \red{Both stages use the same weighted XGBoost
hyperparameter family as the standalone XGB baseline
(Section~\ref{sec:training_xgb_model} and Eq.~\eqref{eq:weight_balance}), with
inverse-frequency class weights computed from the relevant subtraining labels
at each stage.}

Final class probabilities are obtained by multiplicative blending of the
gated-fusion and hierarchical outputs. The hierarchical blend weight is selected
on the validation split from $\{0, 0.15, 0.3, 0.5, 0.75, 1.0\}$. After blending,
class-specific decision thresholds are optimized on the validation set using a
TI-aware objective that combines macro-F1 with an additional topological-class
term and a small accuracy term. The selected blend weight and thresholds are
then applied unchanged to the held-out test set. This procedure yields a
validation-optimized late-fusion model in which semantic and numerical
information interact through both learned gating and hierarchical decision
refinement.

Performance was evaluated on the validation and held-out test sets using the
standard metrics: accuracy, macro-F1, weighted-F1, per-class
precision/recall/F1, and confusion matrix.

\begin{figure}[htbp]
    \centering
    \includegraphics[width=0.95\linewidth]{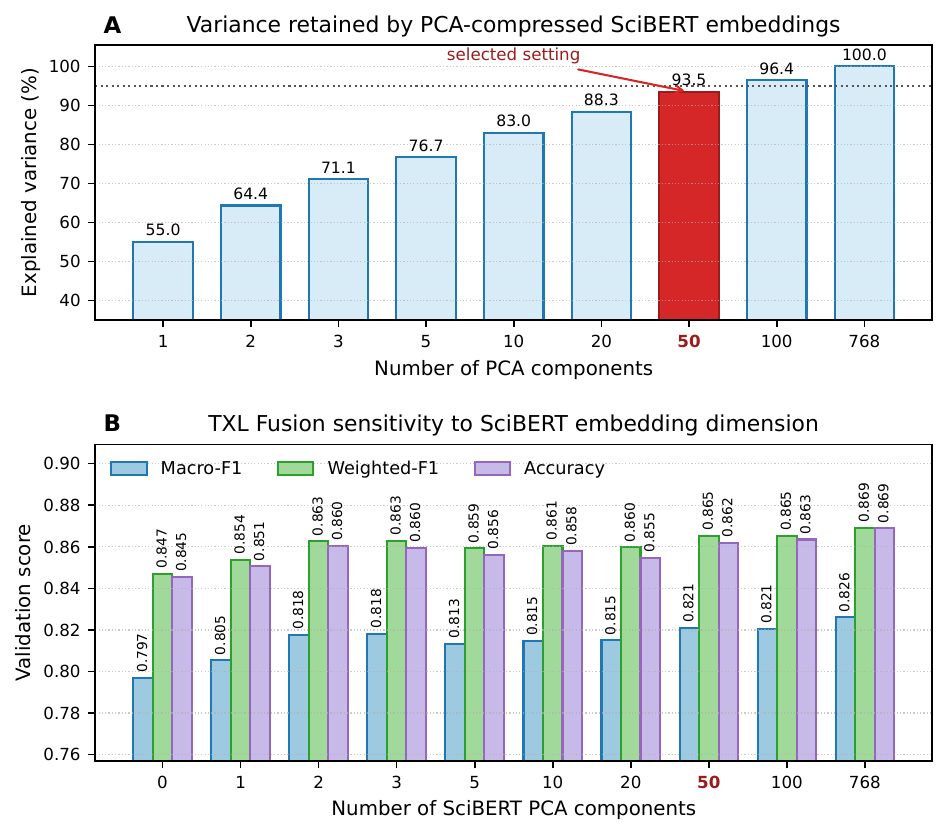}
    \caption{\red{PCA dimensionality analysis for fine‑tuned SciBERT embeddings.
    (A) Cumulative explained variance of the PCA projection fitted on the
    subtraining split. The 50‑component representation retains 93.5\% of the
    variance.
    (B) Validation performance of TXL Fusion as a function of the number of
    retained SciBERT PCA components. The 0‑component model uses only
    numerical/heuristic features; the 768‑component model uses the full
    embedding. Values are printed above each bar.}}
    \label{fig:pca_variance_f1_sensitivity}
\end{figure}

\section{Variance and performance sensitivity of PCA-compressed SciBERT embeddings}
\label{sec:pca_variance}

\red{To justify the dimensionality reduction applied to the semantic branch of TXL Fusion, we quantified the variance retained by PCA on the fine-tuned SciBERT embeddings and examined the sensitivity of model performance to the number of retained components. The analysis was repeated using the corrected semantic narratives, in which SG priors were fitted exclusively on the subtraining split, together with the best weighted SciBERT checkpoint. PCA was fitted only on the 23,644 subtraining embeddings and subsequently applied to the 5,912 validation embeddings, thereby avoiding information leakage.

The cumulative explained variance is shown in Fig.~\ref{fig:pca_variance_f1_sensitivity}A. The first principal component alone captures 55.0\% of the embedding variance, indicating substantial redundancy in the 768-dimensional semantic representation. The first 5 and 10 components capture 76.7\% and 83.0\% of the variance, respectively, while 50 components retain 93.5\% of the total variance. By comparison, 100 components preserve 96.4\%, suggesting that most semantically relevant information is concentrated in a relatively low-dimensional subspace.

We next evaluated the effect of PCA dimensionality on TXL Fusion performance under otherwise identical weighted training settings (Fig.~\ref{fig:pca_variance_f1_sensitivity}B). The handcrafted-feature baseline without semantic embeddings (0 components) achieves a validation macro-F1 of 0.797, weighted-F1 of 0.847, and accuracy of 0.845. Incorporating semantic components consistently improves all metrics, demonstrating that the SciBERT branch contributes complementary predictive information beyond handcrafted descriptors alone. Performance improves rapidly at low dimensionality and then approaches saturation: 10 components already yield substantial gains, while 20, 50, 100, and the full 768-dimensional embeddings produce progressively smaller improvements. At 50 components, performance is already close to the full embedding, achieving a macro-F1 of 0.821, weighted-F1 of 0.865, and accuracy of 0.862, compared with 0.826, 0.869, and 0.869 for the full 768-dimensional representation.

The modest performance gap between 50 components and the full embedding indicates that aggressive compression preserves nearly all task-relevant semantic information. We therefore adopt a 50-component representation for TXL Fusion, as it provides a compact and computationally efficient semantic encoding while retaining most of the predictive benefit of the full SciBERT embedding.}

\section{Semantic versus numerical descriptor: performance comparison} \label{sec:sem-vs-num}

\vspace{0.7cm}  

\begin{figure}[htbp]
\centering
\begin{tcolorbox}[
    enhanced,
    sharp corners,
    colback=lightblue!20,
    colframe=lightblue!50!black,
    title=\textbf{LLM Semantic Descriptor},
    fonttitle=\bfseries,
    drop fuzzy shadow,
    width=\textwidth,
    height=0.35\textheight,
    left=5mm,
    right=5mm,
    top=8mm,
    bottom=5mm
]
\footnotesize\ttfamily

\quad Task: classify the compound as trivial, semimetal, or topological.
\quad Formula: K$_2$Ce$_1$Si$_6$O$_{15}$.
\quad Space\_group: 15.
\quad Space\_group\_prior: trivial = 0.7888, semimetal = 0.1357, topological = 0.0755, support = 914.
\quad Space\_group\_zero\_probability\_classes: none.
\quad Element\_category\_percentages: \{'alkali metal': '8.33', 'alkaline earth metal': '0.00', 'nonmetal': '62.50', 'halogen': '0.00', 'noble gas': '0.00', 'transition metal': '0.00', 'metal': '0.00', 'lanthanide': '4.17', 'actinide': '0.00', 'metalloid': '25.00'\}.
\quad Mean\_valence\_electrons: s = 1.92, p = 3.00, d = 0.04, f = 0.04.
\quad Orbital\_flags: \{'d\_present': True, 'f\_present': True, 'd\_and\_f\_present': True\}.
\quad Total\_electrons: 300.0; parity = even.
\quad Bonding\_nature: moderately ionic.
\quad Topogivity\_scores: trivial\_g = 0.7640, semimetal\_g = -0.6994.
\quad Topogivity\_score\_pattern: \{'trivial\_g\_positive': True, 'semimetal\_g\_positive': False, 'both\_scores\_negative': False, 'opposite\_trivial\_positive\_semimetal\_negative': True, 'opposite\_semimetal\_positive\_trivial\_negative': False\}.
\quad Use the structured evidence above to determine the correct class.

\end{tcolorbox}
\hfill
\begin{tcolorbox}[
    enhanced,
    sharp corners,
    colback=lightgreen!20,
    colframe=lightgreen!50!black,
    title=\textbf{Numerical Descriptor},
    fonttitle=\bfseries,
    drop fuzzy shadow,
    width=\textwidth,
    height=0.30\textheight,
    left=5mm,
    right=5mm,
    top=8mm,
    bottom=5mm
]
\footnotesize\ttfamily
\quad "SG": 15,
\quad "Total\_electrons": 300.0,
\quad "Mean\_s\_valence\_electrons": 1.92,
\quad "Mean\_p\_valence\_electrons": 3.00,
\quad "Mean\_d\_valence\_electrons": 0.04,
\quad "Mean\_f\_valence\_electrons": 0.04,
\quad "Is\_d\_val\_electrons\_present?": 1,
\quad "Is\_f\_val\_electrons\_present?": 1,
\quad "Is\_d\_and\_f\_val\_electrons\_present?": 1,
\quad "Is\_total\_electrons\_even?": 1,
\quad "Ionic\_value": 1.44,
\quad "Bonding\_is\_moderately\_ionic": 1,
\quad "Trivial\_SG\_prob": 0.7888,
\quad "SM\_SG\_prob": 0.1357,
\quad "TI\_SG\_prob": 0.0755,
\quad "SG\_prior\_support": 914,
\quad "Is\_trivial\_SG\_prob\_zero": 0,
\quad "Is\_SM\_SG\_prob\_zero": 0,
\quad "Is\_TI\_SG\_prob\_zero": 0,
\quad "Trivial\_g": 0.7640,
\quad "SM\_g": -0.6994,
\quad "Is\_trivial\_g\_positive": 1,
\quad "Is\_sm\_g\_positive": 0,
\quad "Are\_both\_g\_negative": 0,
\quad "Is\_trivial\_positive\_sm\_negative": 1,
\quad "Is\_sm\_positive\_trivial\_negative": 0,
\quad "label": "trivial",
\quad "Alkali\_metal": 8.33,
\quad "Alkaline\_earth\_metal": 0.0,
\quad "Nonmetal": 62.50,
\quad "Halogen": 0.0,
\quad "Noble\_gas": 0.0,
\quad "Transition\_metal": 0.0,
\quad "Metal": 0.0,
\quad "Lanthanide": 4.17,
\quad "Actinide": 0.0,
\quad "Metalloid": 25.0
\end{tcolorbox}
\caption{\red{Comparative representation of the descriptors for Ce$_1$K$_2$O$_{15}$Si$_6$ (SG~15). The semantic descriptor (top) organizes physical evidence into a structured narrative, while the numerical descriptor (bottom) factorizes the same information into independent scalar features.}}
\label{fig:descriptors-comparison}
\end{figure}

\begin{figure}
    \centering
    \includegraphics[width=\linewidth]{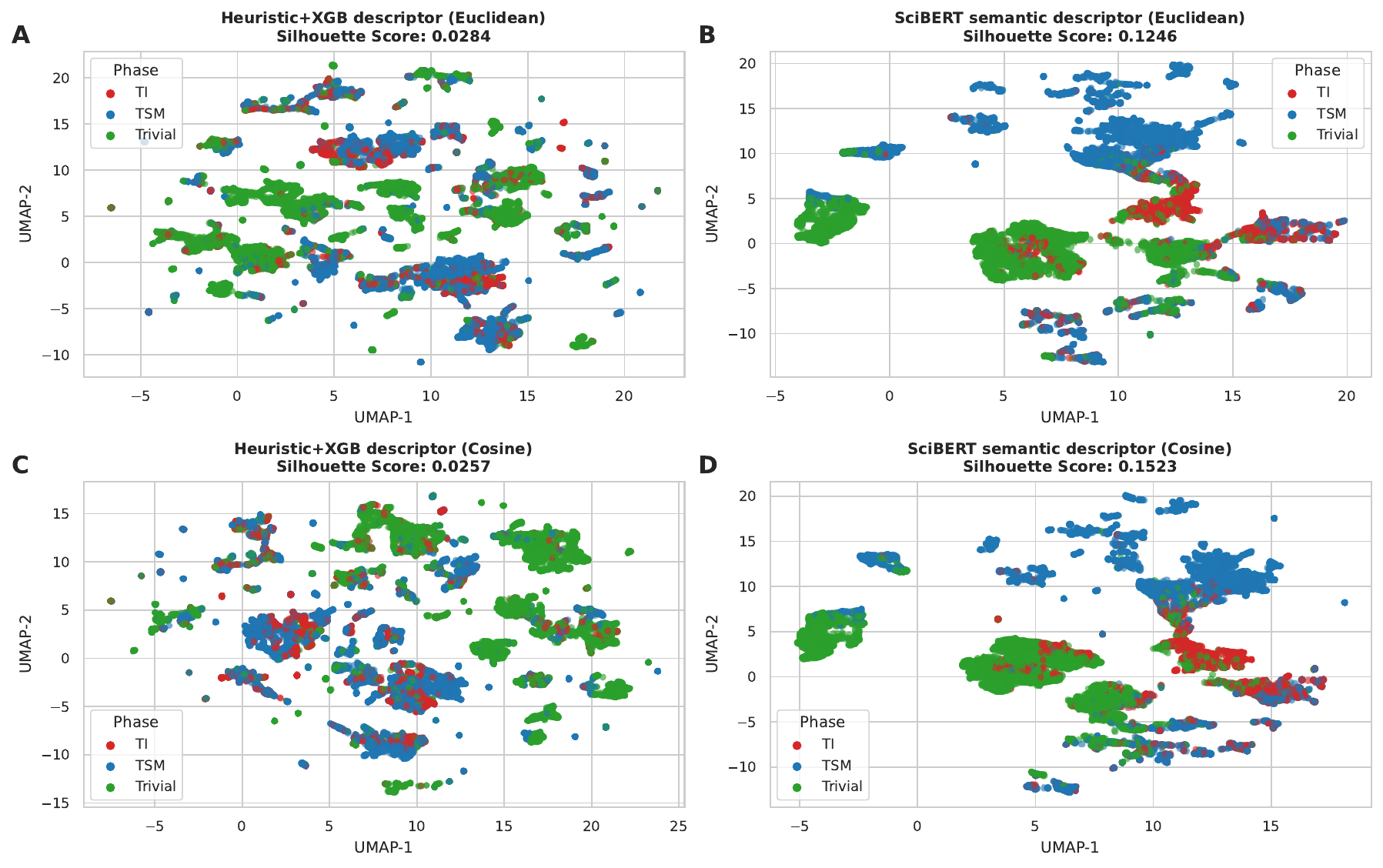}
    \caption{\red{UMAP visualization of topological phase clustering. The low‑dimensional projections compare the weighted heuristic+XGB numerical descriptors (left column: A and C) with the improved SciBERT semantic embeddings (right column: B and D), using Euclidean (top) and cosine (bottom) metrics. Data points are colored by class: trivial (green), TSM (blue), TI (red). Silhouette scores, reported in each panel, quantify cluster separation.}}
    \label{fig:umap}
\end{figure}

Conventional numerical descriptors represent materials through a collection of
independent, hand-crafted features—symmetry labels, electron counts, orbital
occupations, elemental ratios, and heuristic topology indicators. While
interpretable, this factorized formulation cannot explicitly capture coupled
physical effects, such as the interplay between crystal symmetry, orbital
character, and electron filling that governs band inversion and
symmetry-protected states; these correlations must be inferred indirectly by
the downstream model.

Semantic descriptors overcome this limitation by embedding physical knowledge
and domain heuristics within a unified natural-language context.
Figure~\ref{fig:descriptors-comparison} illustrates the contrast for
Ce$_1$K$_2$O$_{15}$Si$_6$ (SG~15). The semantic narrative integrates
compositional ratios, orbital-resolved valence information, symmetry-aware
priors, and qualitative physical reasoning into a single coherent input.
Conditional expert insights—such as the connection between $d/f$-electron
content and topological behavior, or the significance of electron-count
parity—are expressed contextually rather than discretized into binary flags.
When processed by a pretrained language model, this narrative is mapped to a
context-aware embedding that inherently fuses data-driven features with domain
knowledge. In contrast, the numerical descriptor distributes the identical
information across a high-dimensional vector of uncorrelated scalars,
fragmenting the physical relationships that the semantic representation
preserves. Although both descriptors yield a trivial classification for this
example, the semantic formulation retains the inter-dependencies lost in the
factorized numerical feature space.

The effectiveness of the semantic approach is further amplified by transfer
learning. SciBERT is pretrained on an extensive scientific corpus spanning
condensed-matter physics and materials science, and its embedding space inherits
latent physical regularities: correlations between heavy elements and strong
SOC, specific SGs and symmetry-protected degeneracies,
and the orbital fingerprints of non-trivial band topology. Self-attention
mechanisms condition the representation of one concept on the simultaneous
presence of others; for example, elemental identity is encoded jointly with
symmetry context, enabling the embedding to express the potential for
symmetry-protected band crossings or SOC-driven band inversion before
classification. These coupled relationships have no explicit counterpart in
conventional numerical feature spaces.

The improved physical expressiveness is directly reflected in the geometry of
the embedding space. UMAP projections (Fig.~\ref{fig:umap}) reveal that
the heuristic+XGB numerical descriptors yield only weak phase separation, with
silhouette scores of 0.0284 (Euclidean) and 0.0257 (cosine), indicating
substantial overlap among trivial, TSM, and TI compounds. In
contrast, the SciBERT semantic embeddings with only 50 PCA components produce markedly better
clustering, with silhouette scores of 0.1246 (Euclidean) and 0.1523 (cosine).
The consistency of this improvement across distance metrics confirms that the
enhanced separability originates from the intrinsic geometry of the learned
representation rather than from a particular choice of similarity measure.
Semantic descriptors, through LLM embeddings, thus complement
conventional numerical features by capturing the context-dependent
electronic–structural correlations that are essential for discriminating subtle
topological regimes.

\red{
\section{Performance, calibration, and robustness across chemical complexity} \label{sec: el-res-conf-analysis}
\vspace{0.7cm}  

To understand how chemical composition governs model reliability, we evaluate
TXL Fusion and the numerical XGB baseline along three complementary axes:
element‑resolved classification accuracy, prediction confidence and
calibration, and generalization under data sparsity. We also analyze in
detail the origins of reduced performance in chemically simple compounds.

\subsection{Element‑resolved classification performance} \label{sec:elem-resolved-performance}

Performance trends as a function of the number of constituent elements are summarized in Fig.~\ref{fig:element_count_bars}. Across most composition-complexity regimes, TXL Fusion improves held-out performance over the standalone XGB baseline, with the clearest gains observed for the minority TI class. For trivial compounds, TXL maintains strong performance across element-count bins, increasing the unary F1-score from 0.79 to 0.89 and the six-element F1-score from 0.94 to 0.97, while remaining unchanged for five-element compounds (0.97 for both models). For TSM compounds, TXL improves F1 in most bins, increasing performance from 0.75 to 0.79 for unary systems, from 0.83 to 0.86 for binaries, from 0.86 to 0.89 for ternaries, from 0.87 to 0.93 for five-element compounds, and from 0.83 to 0.91 for six-element compounds. The only exception is the four-element TSM bin, where TXL is slightly lower than XGB (0.83 versus 0.84).

The most systematic improvement is observed for TI compounds. TXL improves TI F1 across every element-count bin, increasing performance from 0.29 to 0.39 for unary compounds, from 0.62 to 0.66 for binaries, from 0.66 to 0.73 for ternaries, and from 0.50 to 0.59 for quaternaries. Similar gains are observed in the sparse higher-complexity regimes, where TI F1 increases from 0.44 to 0.50 for five-element compounds and from 0.33 to 0.40 for six-element compounds. Although the six-element TI bin contains only three held-out examples and should therefore be interpreted cautiously, the direction of improvement is consistent across all TI composition-complexity regimes.

Overall, these results indicate that TXL Fusion improves classification robustness across both chemically simple and complex materials. The gains are especially consistent for topological materials, suggesting that semantic--numerical fusion captures additional physicochemical signals needed for reliable TI identification while preserving strong trivial and TSM performance in most element-count regimes.

\begin{figure}
    \centering
    \includegraphics[width=1.0\linewidth]{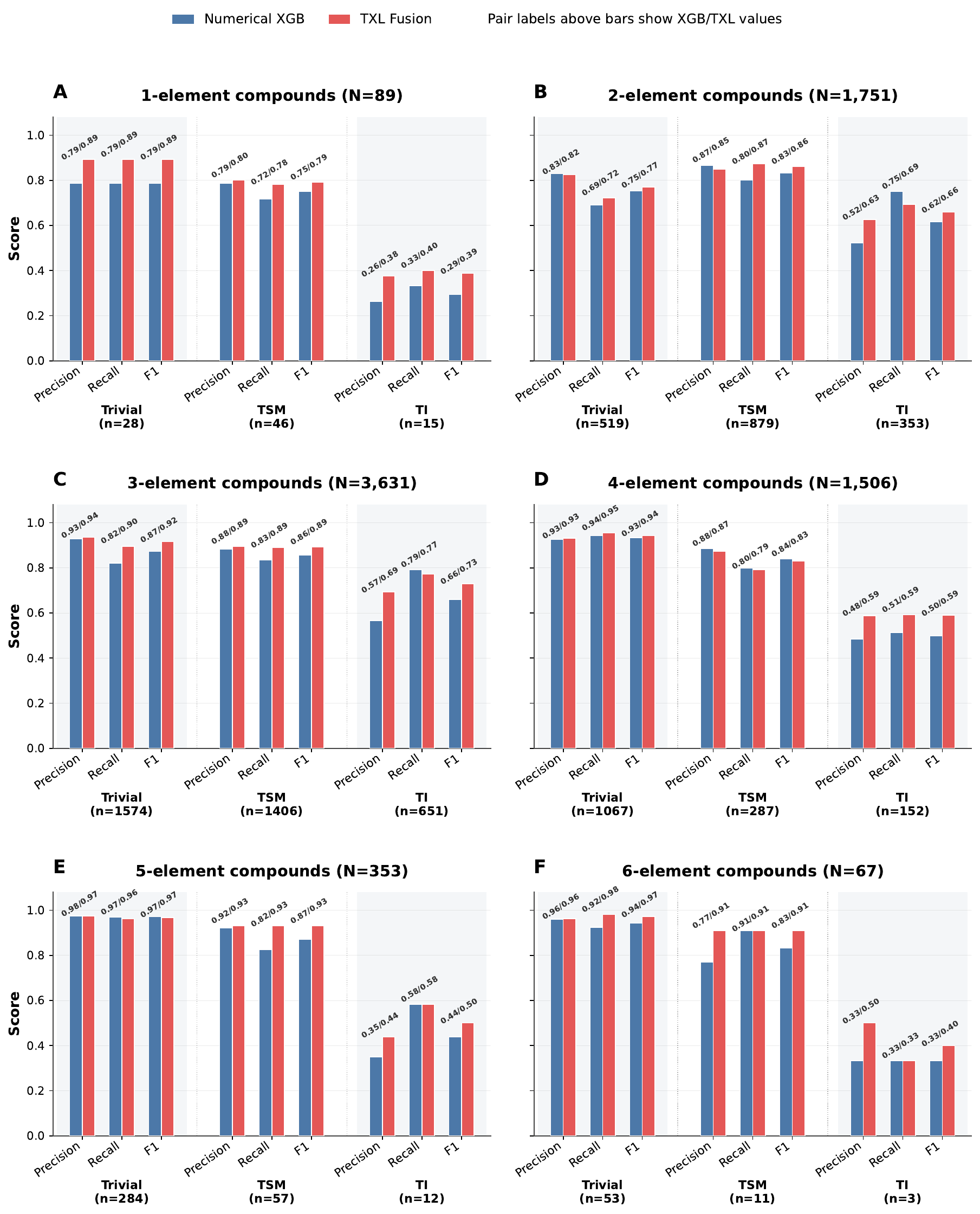}
\caption{\red{Element-resolved comparison of held-out test performance for the standalone XGB model and TXL Fusion. Panels A--F show results for compounds containing one to six constituent elements, respectively. Within each panel, precision, recall, and F1-score are reported for the trivial, TSM, and TI classes, with paired bars comparing XGB and TXL Fusion. Class support is indicated below each class label, and the total number of compounds in each element-count subset is shown in the panel title. Compact labels above each metric pair report the exact XGB/TXL values for that pair.}}
    \label{fig:element_count_bars}
\end{figure}

\subsection{Origins of reduced performance in low-complexity compounds}
\label{sec:low_complexity_performance}
\vspace{0.7cm}

\begin{figure}[htbp]
    \centering
    \includegraphics[width=1.0\linewidth]{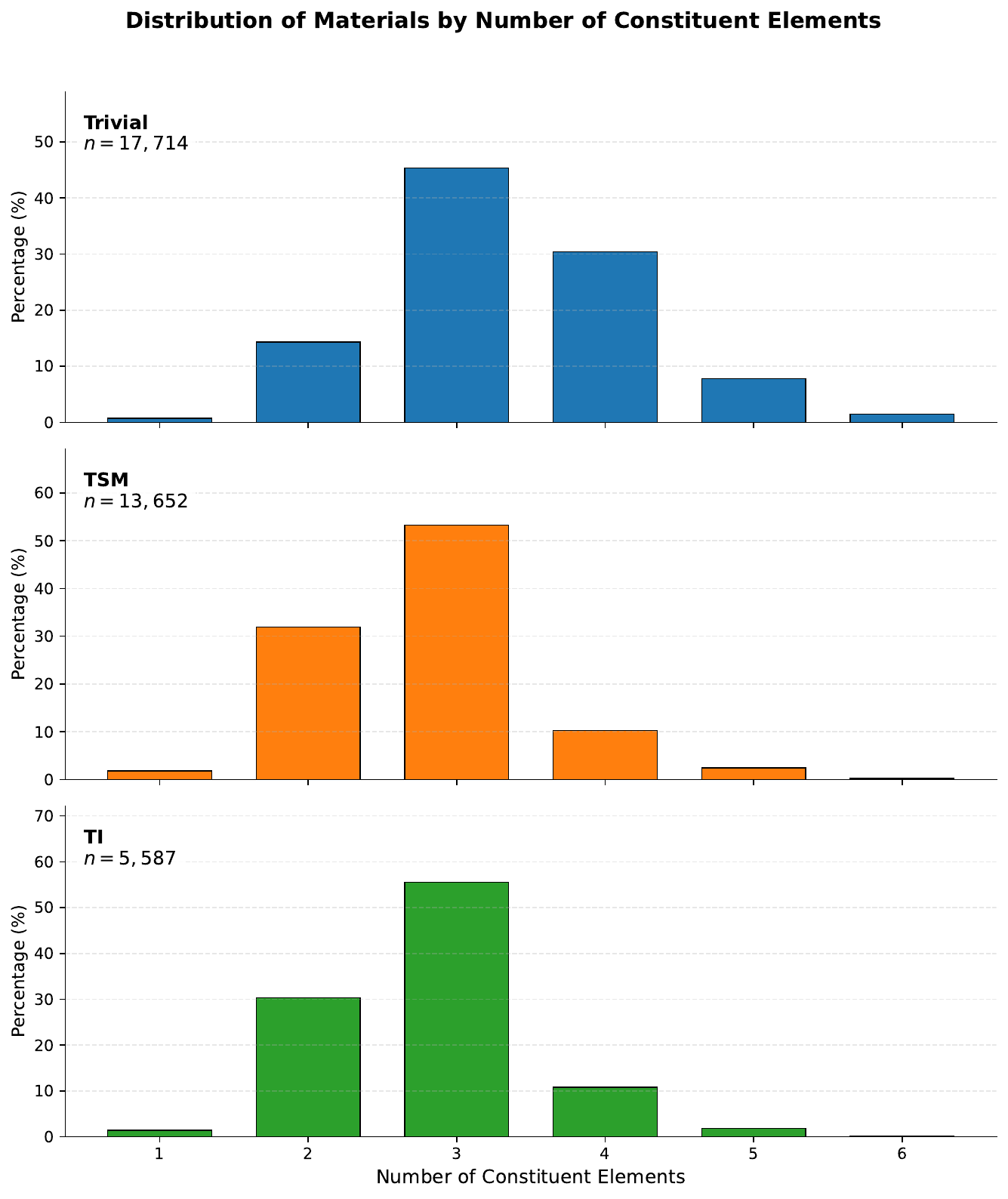}
    \caption{\red{Distribution of compounds by the number of constituent elements across the three topological classes. The dataset is dominated by ternary and quaternary systems, whereas unary and binary compounds are comparatively sparse. This imbalance is particularly pronounced for TIs, for which unary and six-element compounds account for only 1.4\% and 0.13\% of the cleaned dataset, respectively.}}
    \label{fig:el_dist}
\end{figure}

The element-resolved analysis in Fig.~\ref{fig:element_count_bars} shows that predictive performance deteriorates at the chemical extremes, particularly for low-complexity unary and binary compounds. This degradation is consistently observed across trivial, TSM, and TI classes, suggesting that reduced performance in low-complexity materials arises from a combination of statistical and representational limitations.

The first contributing factor is data sparsity. Unary compounds are rare across all classes, with only 127 trivial, 250 TSM, and 79 TI examples, corresponding to 0.72\%, 1.83\%, and 1.41\% of each class, respectively (Fig.~\ref{fig:el_dist}). Although binary compounds are more abundant, their distribution remains strongly class-dependent. By contrast, the dataset is dominated by ternary and quaternary systems, which provide substantially richer training support. This imbalance reduces the statistical coverage of chemically simple materials and makes generalization more difficult at the low-complexity edge of the composition space.

However, sample scarcity alone does not fully explain the observed performance drop. A more fundamental limitation arises from descriptor degeneracy. The numerical descriptors used in this work—including elemental category ratios, electronegativity differences, orbital-resolved valence occupations, and topogivity scores—derive much of their discriminative power from contrasts among multiple atomic species. In unary systems, many of these descriptors collapse to constants or near-constants, while in binary systems they span a substantially narrower range than in ternary, quaternary, or higher-order compounds. As a result, the effective descriptor manifold contracts precisely in the regime where subtle electronic distinctions become most important.

This interpretation is supported by comparing performance across composition-complexity regimes. Six-element compounds are among the rarest groups in the full dataset, yet they often achieve higher F1-scores than unary compounds. For example, six-element trivial and TSM compounds achieve F1-scores of 0.97 and 0.91, respectively, compared with 0.89 and 0.79 for unary compounds. Even for TIs, where both unary and six-element bins are extremely sparse, six-element compounds slightly outperform unary compounds (0.40 versus 0.39). Similarly, the TI binary regime remains challenging despite relatively strong representation: binary TIs account for 30.30\% of the TI subset, compared with only 10.81\% for quaternary TIs, yet binary TIs achieve lower F1 than ternary TIs (0.66 versus 0.73). Together, these observations indicate that low-complexity compounds are difficult not merely because they are under-represented, but because reduced compositional diversity limits descriptor separability.

A third limitation is physical. Certain low-complexity topological materials derive their non-trivial topology from fine electronic-structure effects that are only weakly encoded by composition-level descriptors. Strained $\alpha$-Sn provides a representative example: its topological phase emerges from a symmetry-sensitive band inversion involving Sn-derived states, with SOC coupling and strain controlling whether the system realizes a TI or semimetallic phase.\cite{fu2007topological,barfuss2013elemental,xu2017elemental} For such unary materials, the model must infer band-inversion physics indirectly from SG symmetry, electron filling, orbital character, and SOC-related elemental proxies. The present feature set does not explicitly encode symmetry eigenvalues, irreducible representations, band ordering, orbital-projected band-edge character, or strain-dependent inversion effects.

Future improvements for low-complexity materials will likely require descriptors that directly encode band topology, including symmetry-indicator eigenvalues, orbital-projected band-edge features, or irreducible-representation information at high-symmetry momenta.\cite{bradlyn2017topological,tang2019comprehensive,vergniory2019complete,vergniory2022all} Such descriptors typically require first-principles electronic-structure calculations and are not uniformly available for large-scale screening. The present framework therefore prioritizes rapid composition- and symmetry-based screening to identify candidate families, reserving detailed electronic-structure analysis for targeted follow-up.

\subsection{Generalization of TXL Fusion under chemical sparsity}
\label{sec:txl_chemical_sparsity}
\vspace{0.7cm}

To evaluate whether semantic--numerical fusion improves robustness in poorly
sampled chemical regimes, we compared TXL Fusion with the standalone numerical
XGB model on the sparsest element‑count subsets
(Fig.~\ref{fig:element_count_bars}).
As shown in Fig.~\ref{fig:el_dist}, Unary and six-element compounds are rare across all three classes: they represent only 0.72\% and 1.47\% of trivial materials, 1.83\% and 0.25\% of TSMs, and 1.41\% and 0.13\% of TIs, respectively. Five-element compounds are also sparse for TSMs and TIs, although more frequent among trivial phases.

Despite this limited support, TXL Fusion improves held-out performance in most sparse element-count regimes. In the unary bin, TXL increases F1 from 0.79 to 0.89 for trivial materials, from 0.75 to 0.79 for TSMs, and from 0.29 to 0.39 for TIs. Comparable gains are observed in the six-element bin, where trivial, TSM, and TI F1 improve from 0.94 to 0.97, 0.83 to 0.91, and 0.33 to 0.40, respectively. Five-element TSM and TI performance also increase from 0.87 to 0.93 and from 0.44 to 0.50, while five-element trivial F1 remains essentially unchanged at 0.97. More broadly, TXL improves TI F1 in every element-count subset, including binary (0.62 to 0.66), ternary (0.66 to 0.73), and quaternary compounds (0.50 to 0.59). Although the six-element TI result is based on only three held-out examples and should therefore be interpreted cautiously, the consistent direction of TI improvement across all subsets suggests that TXL Fusion improves generalization of topological decision boundaries under chemical sparsity.

Taken together, these results indicate that TXL Fusion is not only stronger in the densely populated composition regimes but also more robust in low-support regions of chemical space. The semantic branch and numerical descriptors provide complementary information that improves sparse-regime generalization, especially for minority-class topological materials, while preserving high trivial and TSM performance

\subsection{Prediction reliability and calibration} \label{sec:reliability-caliberation}

To assess the prediction reliability of baseline XGb model and TXL Fusion, we performed further assessments using top-label reliability diagrams and the Expected Calibration Error (ECE), shown separately for standalone XGB and TXL Fusion in Figs.~\ref{fig:conf_bins_xgb}
and~\ref{fig:conf_bins_txl}. Using ten equal-width confidence bins on the held-out test set, TXL Fusion achieves higher accuracy than the XGB baseline (0.859 versus 0.820; Results and Discussion section in the main text) and a substantially higher mean top-label confidence (0.948 versus 0.807). Thus, the fused model produces more correct predictions and assigns a larger fraction of samples to high-confidence regimes. However, this increased confidence is accompanied by a larger calibration gap: ECE increases from 0.023 for XGB to 0.096 for TXL Fusion, indicating that TXL is more accurate but also more overconfident in probability space.

This trade-off is most evident in the highest-confidence bin (0.9--1.0). TXL Fusion assigns 6,229 of 7,397 held-out predictions to this bin, with a mean confidence of 0.992 and empirical accuracy of 0.911. By contrast, XGB assigns 3,114 predictions to the same bin, with a mean confidence of 0.955 and empirical accuracy of 0.973. Thus, TXL greatly expands the high-confidence prediction set, but its predicted probabilities are less well calibrated in absolute terms. Because the intermediate-confidence TXL bins contain comparatively few samples (45--432 per occupied bin), local bin-level deviations should be interpreted cautiously. Taken together, these results show that TXL Fusion improves classification accuracy and yields a much larger high-confidence decision regime, while post-hoc calibration may further improve the reliability of its predicted probabilities.

\begin{figure}[htbp]
    \centering
    \includegraphics[width=1.0\linewidth]{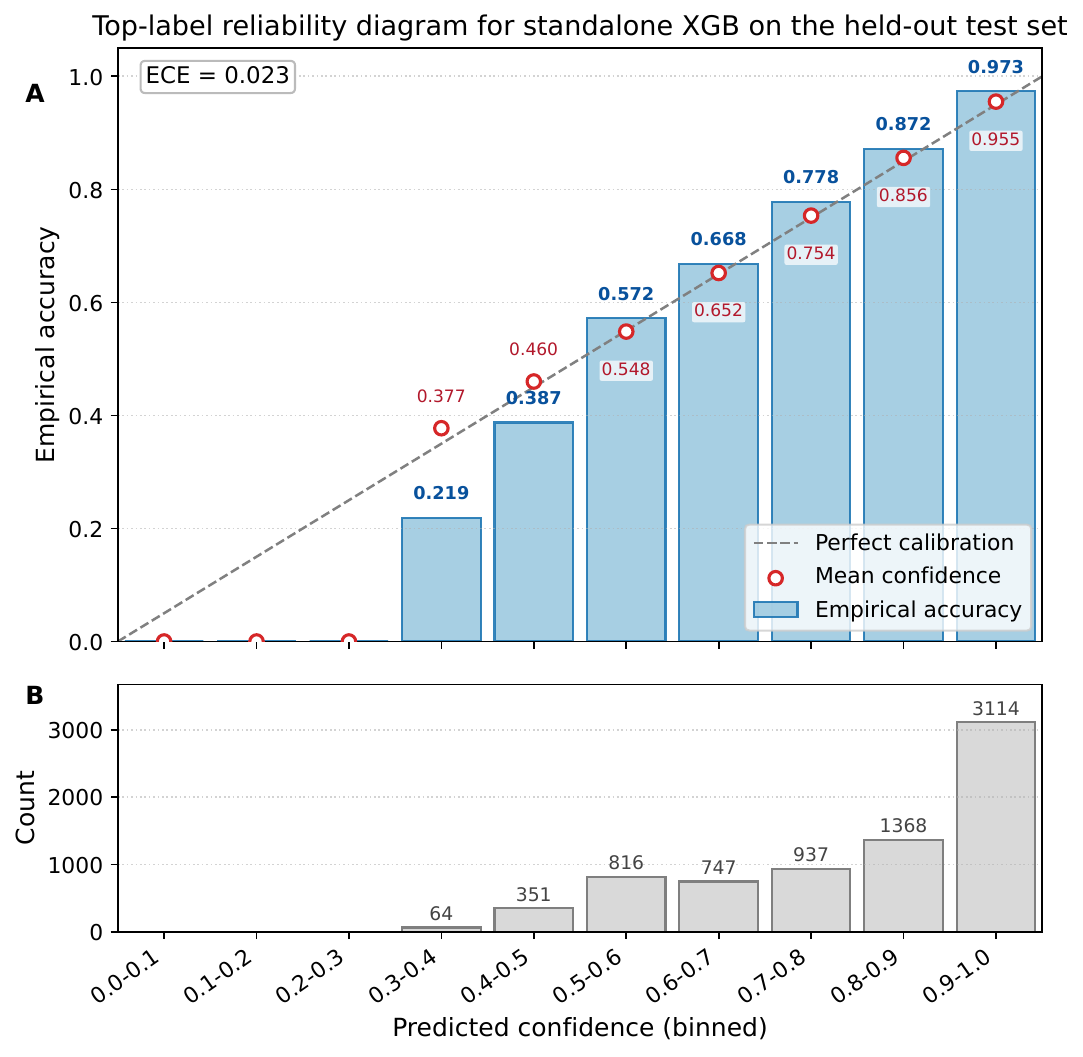}
    \caption{\red{Top-label reliability diagram for the standalone XGB model on the cleaned held-out test set. The upper subplot compares empirical accuracy within each confidence bin against the ideal calibration diagonal and marks the mean confidence of predictions in that bin, while the lower subplot reports the corresponding bin counts. The overall ECE is 0.023.}}
    \label{fig:conf_bins_xgb}
\end{figure}

\begin{figure}[htbp]
    \centering
    \includegraphics[width=1.0\linewidth]{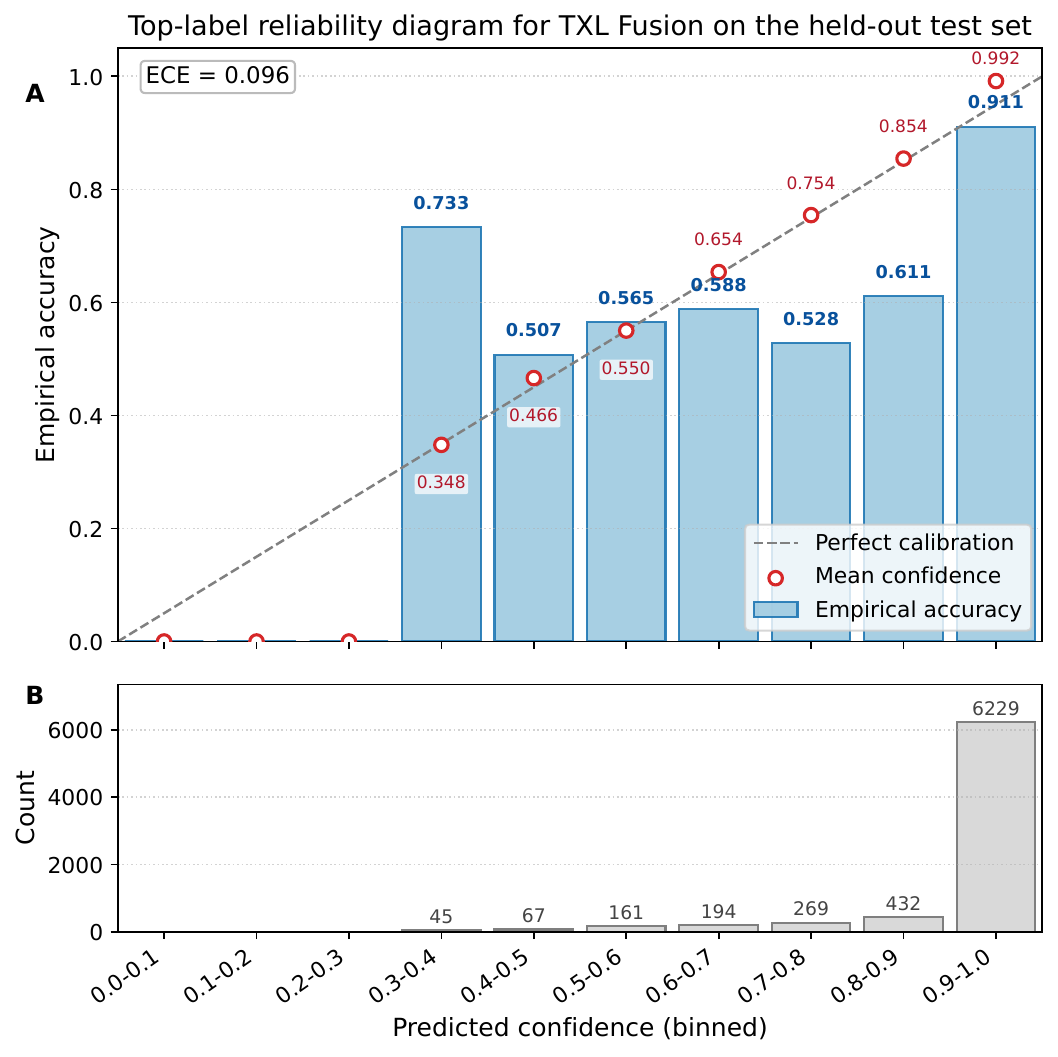}
    \caption{\red{Top-label reliability diagram for TXL Fusion on the cleaned held-out test set. The upper subplot compares empirical accuracy within each confidence bin against the ideal calibration diagonal and marks the mean confidence of predictions in that bin, while the lower subplot reports the corresponding bin counts. The overall ECE is 0.096.}}
    \label{fig:conf_bins_txl}
\end{figure}
}
\red{
\section{Sensitivity of classification to class-imbalance handling} \label{sec:class-imbalance}

\vspace{0.7cm}

To assess the effect of class imbalance, we evaluated standalone XGB and the TXL Fusion XGBoost head under five weighting strategies using the cleaned shared-split protocol. The 29,556-record training pool contains 14,189 trivial, 10,966 TSM, and 4,401 TI compounds, with TI as the minority class (14.9\%). Class weights were computed only from the shared subtraining split (11,351 trivial, 8,772 TSM, and 3,521 TI). The \emph{balanced} setting uses inverse-frequency weights, while the \emph{balanced + TI 1.5$\times$}, \emph{2.0$\times$}, and \emph{3.0$\times$} settings further multiply only the TI weight. All descriptors, hyperparameters, and data splits were kept fixed, and performance was evaluated on the shared validation split and the cleaned 7,397-record held-out test set.

\begin{figure}[htbp]
\centering
\includegraphics[width=1.0\linewidth]{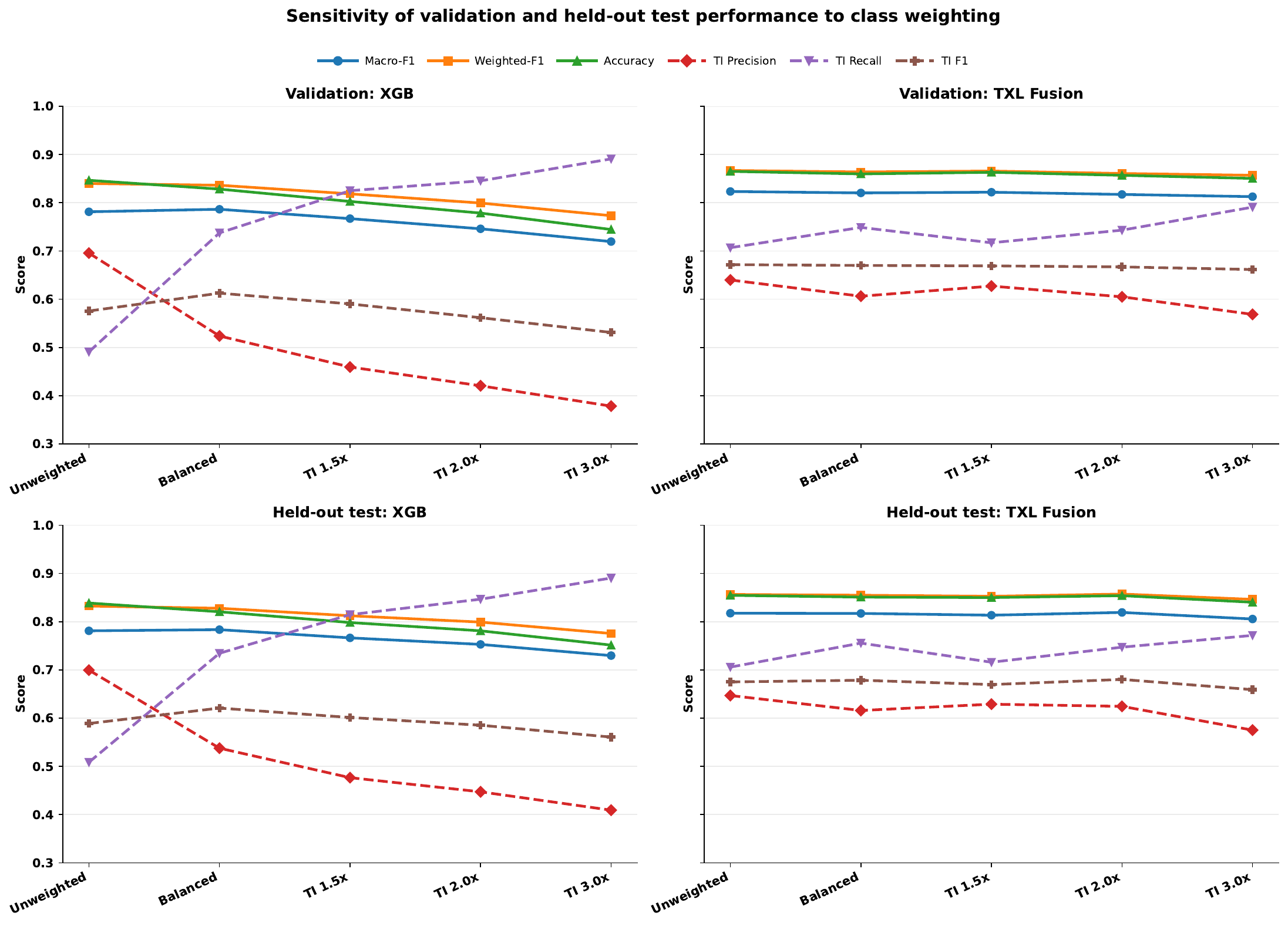}
\caption{\red{Sensitivity of validation and held-out test performance to class-weighting strength under the cleaned shared-split protocol. The four panels show standalone XGB and TXL Fusion on the shared validation split and cleaned held-out test set. Each curve tracks one metric across the five weighting settings: unweighted, \emph{balanced}, \emph{balanced + TI 1.5$\times$}, \emph{balanced + TI 2.0$\times$}, and \emph{balanced + TI 3.0$\times$}. Solid curves show global metrics (Macro-F1, Weighted-F1, and Accuracy), while dashed curves show TI-specific metrics (TI precision, TI recall, and TI F1).}}
\label{fig:class_imbalance_grid}
\end{figure}

Figure~\ref{fig:class_imbalance_grid} shows that standalone XGB is strongly affected by class weighting. Balanced weighting increases TI F1 from 0.5756 to 0.6126 on validation and from 0.5889 to 0.6208 on the held-out test set. However, stronger TI-specific up-weighting progressively lowers TI F1, reaching 0.5312 on validation and 0.5607 on the held-out test set at the 3.0$\times$ setting. Thus, inverse-frequency balancing provides the best TI precision--recall compromise for standalone XGB.

TXL Fusion is markedly more stable across the same sweep. Validation TI F1 changes only modestly from 0.6717 to 0.6701, 0.6691, 0.6670, and 0.6616 from unweighted to 3.0$\times$ TI weighting. Held-out TI F1 remains similarly stable, with values of 0.6750, 0.6785, 0.6696, 0.6802, and 0.6590 across the five settings. Balanced weighting improves held-out TI F1 relative to the unweighted baseline (0.6785 versus 0.6750), while maintaining strong global performance. Although the 2.0$\times$ setting gives the highest held-out TI F1, the gain is small and is not supported by the validation trend.

Global metrics further show the cost of aggressive TI upweighting. For XGB, balanced weighting slightly improves held-out Macro-F1 from 0.7810 to 0.7834, but reduces held-out Weighted-F1 from 0.8323 to 0.8277 and Accuracy from 0.8386 to 0.8205. Stronger TI weighting further decreases held-out Macro-F1 to 0.7664, 0.7528, and 0.7296, and Accuracy to 0.7982, 0.7810, and 0.7514 for the 1.5$\times$, 2.0$\times$, and 3.0$\times$ settings. In contrast, TXL Fusion shows much flatter global behavior, with held-out Macro-F1 remaining within 0.8055--0.8191 and held-out Accuracy within 0.8402--0.8545.

Overall, inverse-frequency balancing offers a robust compromise: it improves TI recovery relative to the unweighted setting without the stronger trade-offs caused by aggressive TI-specific over-weighting. TXL Fusion remains consistently stronger and less sensitive to class-weight selection than standalone XGB, indicating that its combined semantic and descriptor-based representation partly mitigates class-imbalance effects. We therefore adopted the balanced weighting scheme throughout this work.
}
\red{
\section{Robustness to data partitioning}
\label{sec:repeated_split_robustness}
\vspace{0.7cm}

To determine whether the performance advantage of TXL Fusion depends on a particular subtraining--validation partition, we repeated the complete training workflow for both the standalone XGB baseline and the final TXL Fusion model across five independently generated stratified random splits. For each split, the cleaned training pool was repartitioned into 23,644 subtraining records and 5,912 validation records, while the same independent held-out test set of 7,397 compounds was retained for final evaluation. This repeated-partition protocol probes sensitivity to the data partition itself rather than stochastic variation under a fixed split. For TXL Fusion, the full semantic--numerical pipeline was retrained for each partition, including split-specific SciBERT fine-tuning, PCA-based semantic compression, gated fusion, hierarchical refinement, and validation-selected calibration.

\begin{figure*}[htbp]
\centering
\includegraphics[width=0.98\linewidth]{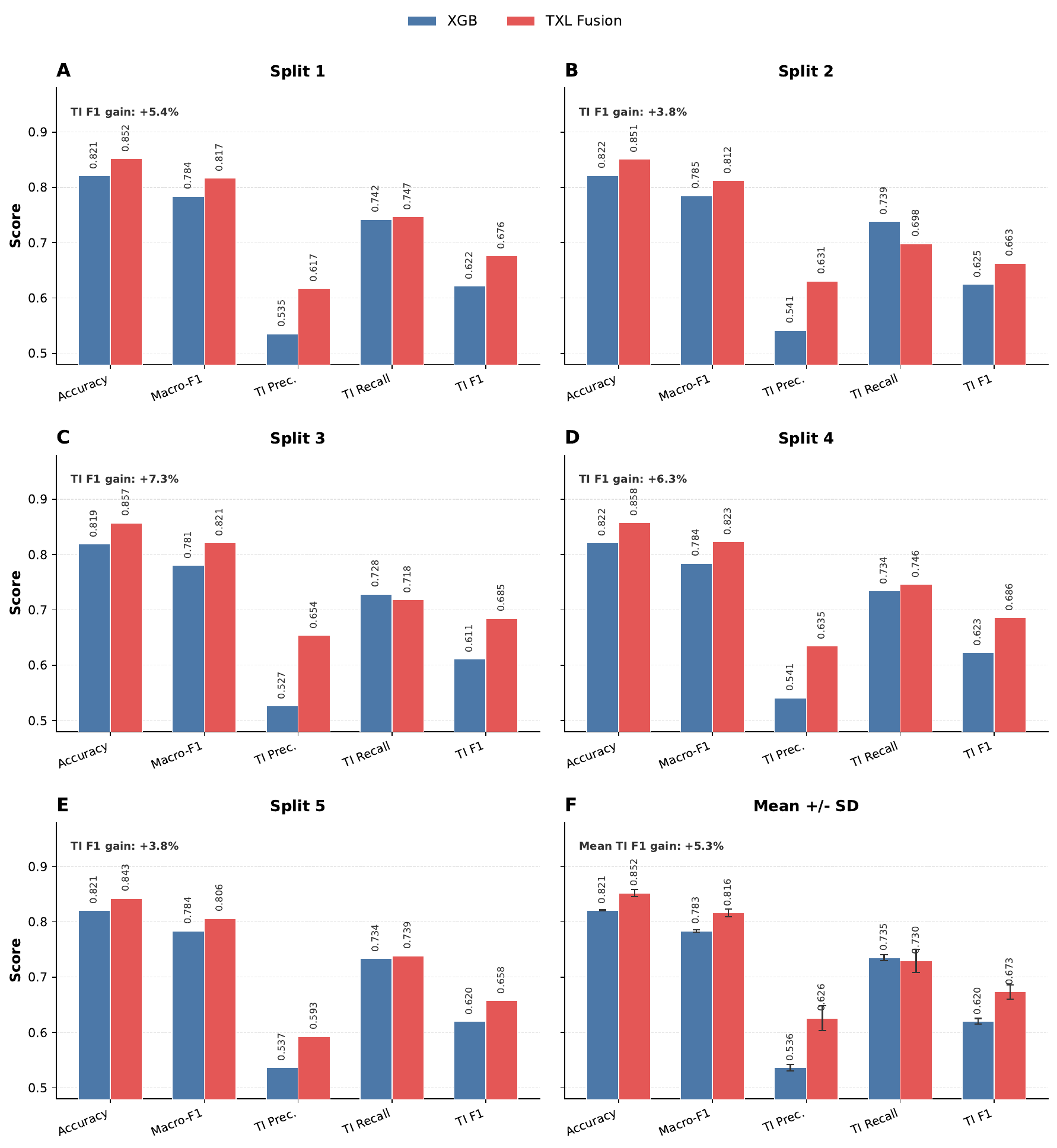}
\caption{\red{Split-wise repeated-partition robustness on the cleaned held-out test set. Panels A--E show held-out performance for five independent random splits, and Panel F reports the mean $\pm$ standard deviation across splits. Bars compare standalone XGB and TXL Fusion using accuracy, macro-F1, TI precision, TI recall, and TI F1. The annotation in each panel indicates the TI F1 gain of TXL Fusion over XGB in percentage points.}}
\label{fig:repeated_split_splitwise_bars}
\end{figure*}

\begin{figure*}[htbp]
\centering
\includegraphics[width=0.94\linewidth]{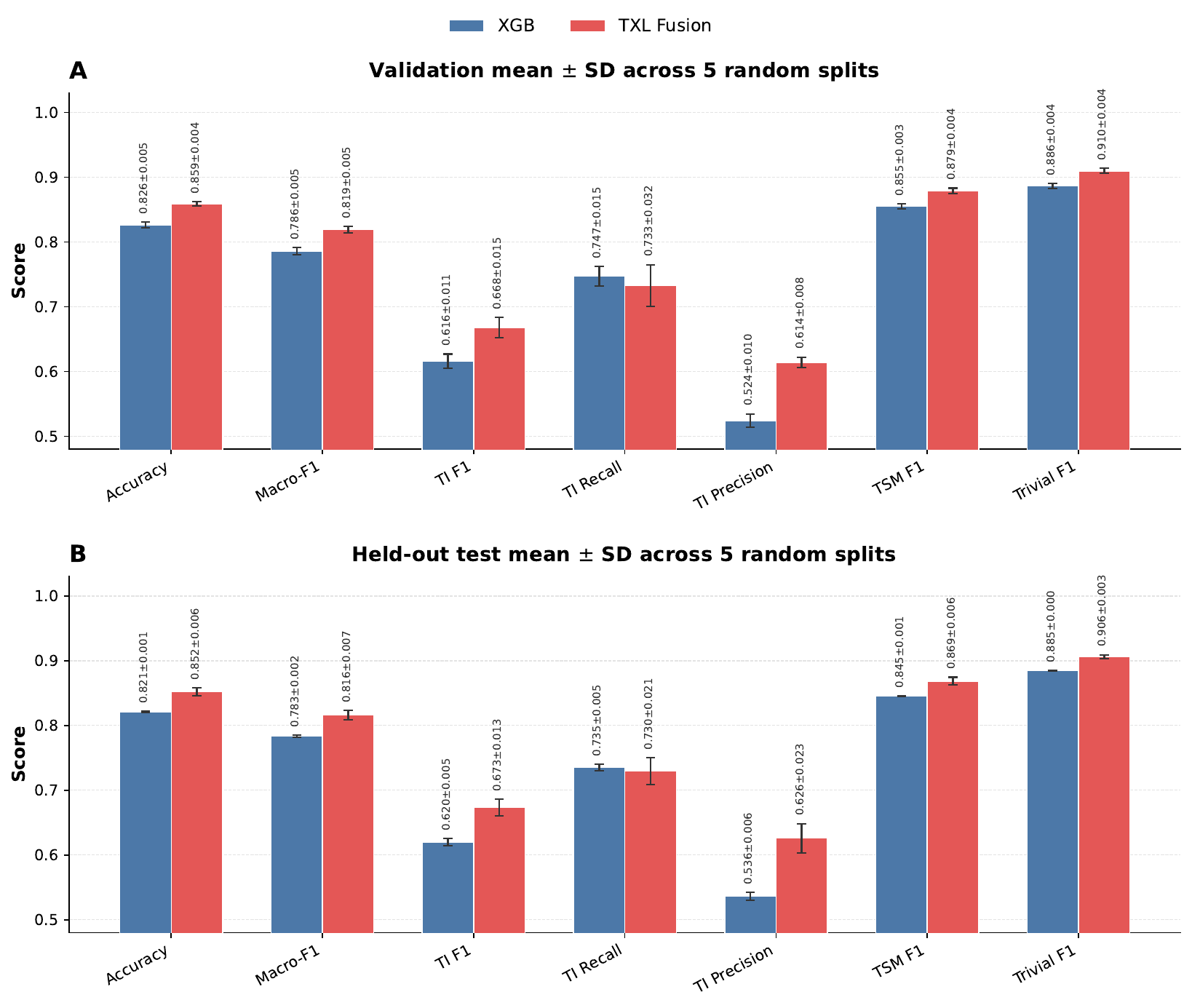}
\caption{\red{Mean repeated-partition robustness of standalone XGB and  TXL Fusion. Panel A summarizes validation performance, and Panel B summarizes performance on the fixed held-out test set. Bars denote the mean across five independently generated random splits, and error bars indicate one standard deviation. Metrics include accuracy, macro-F1, TI F1, TI recall, TI precision, TSM F1, and trivial F1.}}
\label{fig:repeated_split_mean_bars}
\end{figure*}

The split-wise analysis (Fig.~\ref{fig:repeated_split_splitwise_bars}) demonstrates that the TI advantage of  TXL Fusion is reproducible across all five independent partitions. TXL improves TI F1 in every split, with gains of 5.4, 3.8, 7.3, 6.3, and 3.8 percentage points for splits 1--5, respectively, corresponding to a mean improvement of 5.3 percentage points. Averaged across splits, TXL increases TI precision from $0.536 \pm 0.006$ to $0.626 \pm 0.023$, while TI recall remains largely unchanged ($0.735 \pm 0.005$ for XGB versus $0.730 \pm 0.021$ for TXL). Consequently, TI F1 improves from $0.620 \pm 0.005$ to $0.673 \pm 0.013$, indicating that the gain arises primarily from substantially improved precision without sacrificing recall.

This partition robustness extends beyond the TI class (Fig.~\ref{fig:repeated_split_mean_bars}). On the fixed held-out test set,  TXL Fusion improves accuracy from $0.821 \pm 0.001$ to $0.852 \pm 0.006$ and macro-F1 from $0.783 \pm 0.002$ to $0.816 \pm 0.007$. It also improves TSM F1 from $0.845 \pm 0.001$ to $0.869 \pm 0.006$ and trivial F1 from $0.885 \pm 0.000$ to $0.906 \pm 0.003$. Similar gains are observed across validation splits, where accuracy increases from $0.827 \pm 0.005$ to $0.859 \pm 0.004$, macro-F1 from $0.786 \pm 0.005$ to $0.819 \pm 0.005$, and TI F1 from $0.616 \pm 0.011$ to $0.668 \pm 0.015$.

Taken together, these repeated-split experiments show that the semantic--numerical fusion advantage is not an artifact of a favorable partition. Instead, TXL Fusion delivers consistent gains across materially different train--validation decompositions, demonstrating robust generalization and a partition-invariant performance advantage over standalone XGB.

\section{Paired-bootstrap validation of the TXL Fusion advantage}
\label{sec:paired_bootstrap_txl_xgb}
\vspace{0.7cm}

To further assess whether the advantage of TXL Fusion over standalone XGB is robust to finite-sample variation in the held-out set, we performed a paired-bootstrap analysis for each of the five independently generated training--validation partitions described above. In each split, both models were evaluated on the same cleaned held-out test set of 7,397 materials. Held-out compounds were then resampled with replacement using a common bootstrap index set for both models, thereby preserving the compound-wise pairing of XGB and TXL predictions. For each bootstrap replicate, we computed the paired performance difference
\begin{equation}
\Delta =
\mathrm{metric}(\mathrm{TXL}) -
\mathrm{metric}(\mathrm{XGB}).
\end{equation}
We generated 3000 paired bootstrap replicates per split. Table~\ref{tab:bootstrap_txl_xgb} reports the mean observed difference across the five split-level analyses, the corresponding mean 95\% percentile-bootstrap confidence interval, and the mean bootstrap probability that $\Delta>0$.

\begin{table}[htbp]
\centering
\caption{\red{Repeated-split paired-bootstrap comparison of TXL Fusion and standalone XGB on the cleaned held-out test set. Values are reported as TXL Fusion minus XGB under the weighted training protocol. The table gives the mean observed difference across five split-level bootstrap analyses, the mean 95\% percentile-bootstrap confidence interval, and the mean bootstrap probability that the paired difference is positive.}}
\label{tab:bootstrap_txl_xgb}
\small
\begin{tabular}{lccc}
\toprule
Metric & Mean observed $\Delta$ & Mean 95\% CI & Mean $P(\Delta > 0)$ \\
\midrule
\multicolumn{4}{l}{\textbf{Precision}} \\
Trivial precision & $+0.0026$ & $[-0.0049,,0.0104]$ & $0.7110$ \\
TSM precision     & $+0.0049$ & $[-0.0050,,0.0151]$ & $0.7268$ \\
TI precision      & $+0.0898$ & $[0.0719,,0.1078]$ & $1.0000$ \\
\midrule
\multicolumn{4}{l}{\textbf{Recall}} \\
Trivial recall & $+0.0376$ & $[0.0278,,0.0470]$ & $1.0000$ \\
TSM recall     & $+0.0400$ & $[0.0283,,0.0515]$ & $0.9999$ \\
TI recall      & $-0.0057$ & $[-0.0282,,0.0168]$ & $0.4753$ \\
\midrule
\multicolumn{4}{l}{\textbf{F1-score}} \\
Trivial F1 & $+0.0211$ & $[0.0145,,0.0277]$ & $1.0000$ \\
TSM F1     & $+0.0233$ & $[0.0148,,0.0316]$ & $0.9999$ \\
TI F1      & $+0.0533$ & $[0.0364,,0.0703]$ & $1.0000$ \\
Macro-F1   & $+0.0325$ & $[0.0240,,0.0410]$ & $1.0000$ \\
\bottomrule
\end{tabular}
\end{table}

The paired-bootstrap results provide strong evidence that the TXL Fusion improvement is not driven by a small subset of held-out compounds. The largest class-specific gain occurs for TI precision, which increases by (+0.0898), with a strictly positive mean 95\% confidence interval of ([0.0719,,0.1078]) and $P(\Delta>0)=1.0000$. In contrast, TI recall is statistically indistinguishable between the two models on average $(\Delta=-0.0057$, 95\% CI ([-0.0282,,0.0168]), $P(\Delta>0)=0.4753$). Thus, TXL Fusion does not materially alter the fraction of true TI compounds recovered. Rather, the semantic--numerical fusion primarily improves TI discrimination by reducing false-positive TI assignments. This precision-driven improvement translates into a robust TI F1 gain of (+0.0533), with a strictly positive confidence interval of ([0.0364,,0.0703]).

The advantage is not confined to the TI class. TXL Fusion also improves trivial F1 by (+0.0211) and TSM F1 by (+0.0233), with strictly positive confidence intervals for both classes. At the global level, Macro-F1 increases by (+0.0325), with a mean 95\% confidence interval of ([0.0240,,0.0410]), and the bootstrap probability that TXL outperforms XGB is effectively one. Together with the repeated-partition analysis, these paired-bootstrap results demonstrate that the TXL Fusion advantage is stable under both independent data repartitioning and compound-wise resampling of the held-out test set. Importantly, TXL does not simply predict more topological materials; it improves the reliability of TI assignments while simultaneously enhancing overall three-class classification performance. 

\section{DFT computational details}
\label{sec:dft_details}
\vspace{0.7cm}

Density-functional-theory (DFT) calculations were performed using the Vienna \textit{ab initio} Simulation Package (VASP).\cite{kresse1993ab,kresse1996efficiency} Core--valence interactions were described using the projector-augmented-wave (PAW) method,\cite{kresse1999from} and exchange--correlation effects were treated within the Perdew--Burke--Ernzerhof generalized-gradient approximation (PBE-GGA). For each calculation, the order of PAW datasets in the POTCAR file was checked against the species order in the corresponding POSCAR. The plane-wave cutoff was chosen on a material-specific basis from the production INCAR files, typically 400 or 520~eV. When automated cutoff convergence was performed, cutoffs from 300 to 600~eV were tested in 50~eV increments, and the first value for which the total-energy change relative to the preceding cutoff was below 1~meV~atom$^{-1}$ was selected.

The crystal structures associated with the target database entries and space groups were used as starting geometries. In the production topology workflow, internal atomic coordinates were relaxed while retaining the database lattice vectors, using \texttt{IBRION = 2}, \texttt{ISIF = 2}, and \texttt{NSW = 200}. Full cell relaxation (\texttt{ISIF = 3}) was used only for separately designated structure-relaxation tasks. Electronic self-consistency was converged to $10^{-5}$~eV, and ionic relaxation was stopped when the maximum residual force was below 0.015~eV~\AA$^{-1}$ (\texttt{EDIFFG = -0.015}). After relaxation, the composition, atomic ordering, and space group were rechecked before assigning the electronic and topological results to the corresponding database entry.

Structural relaxations used automatic $\Gamma$-centred $k$-point meshes, with the initial number of points along lattice direction $i$ set to $N_i=\max[1,\mathrm{round}(50~\text{\AA}/a_i)]$. Static self-consistent-field (SCF) calculations used meshes regenerated from the relaxed structures with a reciprocal-space resolution of approximately 0.04~\AA$^{-1}$. Gaussian smearing was used in all production electronic-structure calculations (\texttt{ISMEAR = 0}, \texttt{SIGMA = 0.05}~eV). Fractional occupations were retained as diagnostics, but were not used alone to distinguish metallic and insulating states.

Following relaxation, static noncollinear SOC calculations were performed using \texttt{LSORBIT = .TRUE.} and \texttt{LNONCOLLINEAR = .TRUE.}. The SOC SCF calculations used \texttt{ALGO = Normal}, an electronic tolerance of $10^{-5}$~eV, and wrote the converged charge density for subsequent band-structure and Wannier calculations. Nonmagnetic initial states used three zero Cartesian magnetic-moment components per atom. For systems that developed a sizable magnetic moment, or contained plausible magnetic ions, nonmagnetic, ferromagnetic, and representative antiferromagnetic initial states were compared using the same structure and numerical settings. The lowest-energy converged state was retained. Ordinary time-reversal $\mathbb{Z}_2$ indices were not assigned to confirmed magnetic ground states.

SOC band structures were calculated non-self-consistently from the converged SOC charge density using \texttt{ICHARG = 11} and \texttt{ISTART = 0}, along symmetry-derived high-symmetry paths. The fixed electron-count boundary was used to identify the valence-band maximum and conduction-band minimum. A material was classified as insulating only when the indirect gap was positive. Systems with a positive direct gap but negative indirect gap were labelled as indirect-overlap metals, whereas systems with vanishing direct gaps were treated as band-crossing candidates. The numerical tolerance for metal--insulator classification was $10^{-4}$~eV. Small-gap and band-overlap cases were further checked using explicit SOC band paths and, where necessary, denser reciprocal-space sampling.

After SOC band analysis, materials were assigned to separate insulating or metallic/semimetallic post-processing workflows. Insulating systems were further classified using Wilson-loop $\mathbb{Z}_2$ invariants as strong TIs, weak TIs, or trivial insulators. Metallic, indirect-overlap, and band-crossing systems were treated as semimetallic candidates and analysed using Wannier-based node searches, enclosing-surface Chern numbers, Berry phases, and symmetry/connectivity checks. Thus, trivial, TI, and TSM assignments followed distinct topological criteria after the common SOC-DFT and Wannier-validation workflow.

Wannier tight-binding Hamiltonians were constructed for both insulating and metallic systems using the VASP--Wannier90 interface and Wannier90.\cite{mostofi2008wannier90} The VASP Wannier step used a uniform Monkhorst--Pack mesh, \texttt{ICHARG = 11}, and \texttt{NCORE = 1}. Initial projections were generated using the selected-columns-of-the-density-matrix (SCDM) procedure. For insulators, the Wannier model included the complete occupied DFT subspace together with a nonzero number of conduction bands; occupied-only Hamiltonians with \texttt{NumOccupied = num\_wann} were not used for $\mathbb{Z}_2$ analysis. For metals and band-crossing candidates, the Wannier subspace was centred on the valence--conduction frontier near the Fermi level. Outer and frozen disentanglement windows were generated from the SOC SCF and SOC band eigenvalues, and the number of frozen states was checked at each sampled $k$ point. Up to 5000 disentanglement iterations were allowed.

All Wannier models were subjected to fail-closed validation before topological post-processing. This validation checked the SOC/spinor setting, Hamiltonian dimension, occupied-band boundary, energy windows, SCF provenance and convergence, Wannier90 completion, and agreement between DFT and Wannier-interpolated bands. The maximum interpolation error was required to be no larger than 0.15~eV. For metallic and node-search models, the maximum error within 0.2~eV of the Fermi-level frontier was additionally required to be no larger than 0.01~eV. Wannier bands were also compared against an independent line-mode SOC band structure, especially when a reduced Wannier $k$-mesh was used. Results from failed models, as well as node files inherited from earlier or failed jobs, were discarded.

For nonmagnetic insulators with a complete occupied Wannier subspace, the six time-reversal-invariant momentum planes, $k_i=0$ and $k_i=\pi$ $(i=1,2,3)$, were analysed using hybrid-Wannier-charge-centre Wilson loops through the \texttt{Z2\_3D} implementation in WannierTools.\cite{wu2018wanniertools} Z2Pack was used as an independent or fallback implementation when a complete calculation was available.\cite{gresch2017z2pack} The two-dimensional $\mathbb{Z}_2$ index was first obtained on each time-reversal-invariant plane. For each reciprocal direction, the strong index was then computed as the modulo-two sum of the two corresponding parallel-plane indices; the three independently obtained values of the strong index were required to agree as a consistency check. The weak indices were taken from the Wilson-loop indices on the $k_x=\pi$, $k_y=\pi$, and $k_z=\pi$ planes. Materials were classified as strong topological insulators when $\nu_0=1$, weak topological insulators when $\nu_0=0$ with at least one nonzero weak index, and trivial insulators only when all four $\mathbb{Z}_2$ indices were zero.

For metallic and semimetallic candidates, WannierTools \texttt{FindNodes} calculations were performed for all relevant Fermi-frontier band boundaries, rather than assuming a single fixed \texttt{NumOccupied}. Candidate nodes were retained only when $|E-E_F|\leq0.2$~eV and the residual band gap was below 1~meV; periodic duplicates were removed in reduced reciprocal coordinates. A \texttt{FindNodes} minimum was treated only as a candidate degeneracy. A Weyl point was assigned only when a nonzero enclosing-surface Chern number was reproduced for several sphere radii (0.002, 0.004, and 0.006~\AA$^{-1}$). A zero Chern number alone was not used to identify Dirac points or nodal lines. Dirac and nodal-line assignments additionally required the appropriate degeneracy, symmetry protection, momentum-space connectivity, and, for nodal lines, a nontrivial Berry phase on a closed loop linking the proposed line.

For confirmed magnetic metals, the time-reversal $\mathbb{Z}_2$ workflow was not applied. Their topology was instead analysed from the Wannier Hamiltonian using node searches, Chern-number calculations, and, where appropriate, Berry-curvature and anomalous-Hall-conductivity calculations.

\section{Reliability of DFT-based topological labels} \label{sec:labels-reliability}

First-principles calculations underpin large-scale topological materials databases,
but their reliability is not uniform. The database employed in this work
explicitly flags more than 2,300 compounds with partially filled $f$ shells,
where standard DFT is challenged by strong correlations~\cite{vergniory2019complete}.
Beyond this, the dataset contains many contradictory labels for the same
formula and SG combination. To characterize this internal inconsistency,
we performed a direct audit of the original 38,184‑record dataset (before
cleaning) using chemical formula and SG as the only structural
identity available to the model.

Grouping entries by (formula, SG) reveals how often the same coarse key receives
different topological class labels (Table~\ref{tab:label_consistency_audit}).
For a model that relies solely on composition‑ and SG‑derived descriptors, such
cases constitute contradictory supervision, regardless of finer structural
differences that are not represented in the feature set. The audit shows that
575 formula+SG keys carry conflicting labels, encompassing 1,231 entries.
Conflicts involving the TI class are particularly
frequent: 463 keys, with TI vs trivial (254 keys) and TSM vs TI (199 keys) the
most common patterns. A representative example is Cr$_2$O$_3$ in SG~167, which
appears with all three labels across different ICSD entries. Notably, 184
conflicting keys span both the training and held‑out test partitions,
demonstrating that this ambiguity is not confined to a single subset. These
internal contradictions directly motivated the cleaning procedure adopted for
all weighted ablation and TXL experiments.

We emphasize that conflicting entries are not necessarily database errors:
compounds with the same composition and SG can differ in internal
coordinates, magnetic ordering, or other details not encoded in our descriptors.
Nevertheless, for a composition+SG learning task they represent effective label
noise, with the TI class particularly affected.

Even after removing explicit internal contradictions, residual label
disagreement persists across independent data sources. To assess this, we
compared the cleaned full dataset with the Topomat collection~\cite{topomat_dataset} after merging
TSM and TI into a single topological class. Matching was performed by
normalised formula and SG number, and Topomat entries with uncertain topological
invariants ($\nu = ?$) were excluded (702 materials). Among the 9,986 shared
keys, 8,264 showed clean agreement, while 1,624 keys (16.3\%) exhibited a direct
contradiction between the cleaned dataset and Topomat
(Table~\ref{tab:topomat_consistency_audit}). The asymmetry is striking: 1,217
keys are labelled topological in the cleaned dataset but trivial in Topomat,
whereas only 407 are labelled trivial in the cleaned set but topological in
Topomat.

\begin{table}[h!]
\centering
\caption{\red{Internal label-consistency audit of the original 38,184-record dataset
before cleaning. Entries are grouped by normalised formula and SG
number.}}
\label{tab:label_consistency_audit}
\begin{tabular}{lc}
\hline
Statistic & Value \\
\hline
Conflicting formula+SG keys & 575 \\
Entries contained in conflicting keys & 1,231 \\
Conflicting keys involving TI & 463 \\
TI vs trivial conflicts & 254 \\
SM vs TI conflicts & 199 \\
SM vs trivial conflicts & 112 \\
All three labels present for the same formula+SG key & 10 \\
Conflicting keys split across train and held-out test splits & 184 \\
\hline
\end{tabular}
\end{table}

\begin{table}[h!]
\centering
\caption{\red{External consistency audit between the cleaned full dataset and
Topomat. TSM and TI are merged into a single topological group; Topomat entries
with $\nu = ?$ are excluded.}}
\label{tab:topomat_consistency_audit}
\begin{tabular}{lc}
\hline
Statistic & Value \\
\hline
Shared formula+SG keys after excluding unclear Topomat labels & 9,986 \\
Clean single-label agreements & 8,264 \\
Direct cleaned-vs-Topomat contradictions & 1,624 \\
Contradiction rate & 16.3\% \\
Cleaned topological vs Topomat trivial & 1,217 \\
Cleaned trivial vs Topomat topological & 407 \\
Excluded Topomat entries with $\nu = ?$ & 702 \\
\hline
\end{tabular}
\end{table}

It is worth mentioning that we do not regard Topomat as an unquestionable ground truth; rather, this
external comparison shows that even after internal cleaning, a substantial
fraction of overlapping materials receive opposite coarse labels from different
reputable sources. The persistent cross-database inconsistency reinforces the
interpretation that residual label uncertainty remains present in the cleaned
training corpus. Cleaning eliminates the most blatant contradictions, but it
does not resolve the deeper ambiguity inherent to DFT-derived topological
annotations, particularly near the trivial–nontrivial boundary. This intrinsic
label noise likely contributes to the performance ceiling observed for all
models, especially for the TI class.
}
}
\clearpage
\bibliography{references}